\documentclass[12pt]{iopart}
\usepackage{graphicx}
\newcommand{\beq}{\begin{eqnarray}}
\newcommand{\eeq}{\end{eqnarray}}
\begin{document}
\title[Interplay between excitation kinetics and reaction-center dynamics
in purple bacteria] {Interplay between excitation kinetics and
reaction-center dynamics in purple bacteria}

\author {Felipe Caycedo-Soler, Ferney J. Rodr\'{\i}guez and Luis Quiroga}
\address{Departamento de F\'{\i}sica, Universidad de Los Andes, A.A. 4976 Bogot\'a,
D.C., Colombia}
\author {Neil F. Johnson}
\address{Department of Physics, University of Miami, Coral Gables, Miami, Florida 33126, USA}
\ead{f-cayced@uniandes.edu.co}
\date{\today}

\begin{abstract}
Photosynthesis is arguably the fundamental process of Life, since it enables energy from the Sun to enter the food-chain on Earth. It is a remarkable non-equilibrium process in which photons are converted to many-body excitations which traverse a complex biomolecular membrane, getting captured and fueling chemical reactions within a reaction-center in order to produce nutrients. The precise nature of these dynamical processes -- which lie at the interface between quantum and classical behaviour, and involve both noise and coordination -- are still being explored. Here we focus on a striking recent empirical finding concerning an illumination-driven transition in the biomolecular membrane architecture of {\it Rsp. Photometricum} purple bacteria. Using stochastic realisations to describe a hopping rate model for excitation transfer, we show numerically and analytically that this surprising shift in preferred architectures can be traced to the interplay between the excitation kinetics and the reaction center dynamics. The net effect is that the bacteria  profit from efficient metabolism at low illumination intensities while using dissipation to avoid an oversupply of energy at high illumination intensities.
\end{abstract}
\maketitle

\section{Introduction}
In addition to its intrinsic interest as one of Nature's oldest and most important processes,
photo-energy conversion is of great practical interest given Society's pressing need to reduce
reliance on fossil fuels by exploiting alternative energy production. Photosynthesis maintains
the planet's oxygen and carbon cycles in equilibrium \cite{sturg1,oxy,carbon} and efficiently
converts sunlight \cite{effic,pullerits, flemming}, while the possibility of
its {\it in vivo} study provides a fascinating window into the aggregate effect of
millions of years of natural selection. Among the most widespread photosynthetic
systems are purple bacteria {\it Rsp. Photometricum} which manage to sustain
their metabolism even under dim light conditions within ponds, lagoons and
streams  \cite{Pfenning}.  They absorb light through antenna structures in the biomolecular
Light Harvesting complex 2 (LH2), and transfer the  electronic excitation along
the membrane to Light Harvesting complexes 1 (LH1) which each contain a Reaction
Center (RC) complex.  If charge carriers are available (i.e. the RC is in an open state),
then  the resulting reactions will feed the bacterial metabolism.

\begin{figure}
\centering
\includegraphics[width=0.8\columnwidth]{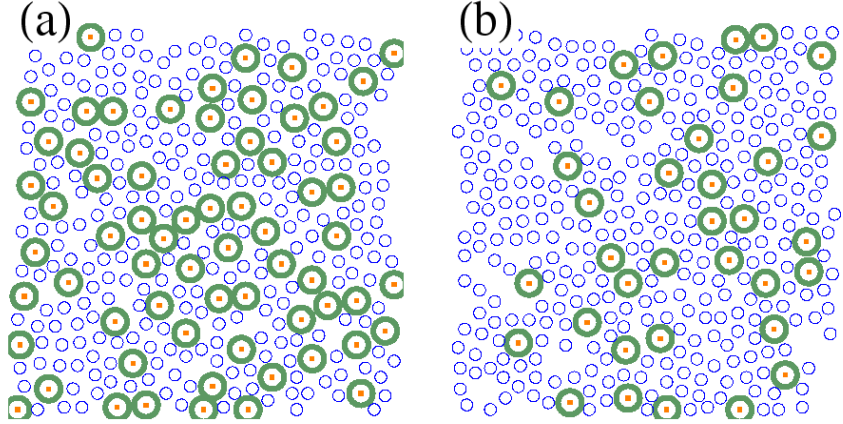}
\caption{Empirical architectures for (a) HLI  and (b) LLI
membranes, displaying LH2s (small blue circles), LH1s (big green
circles) and RCs (orange dots).}\label{membs}
\end{figure}

It was recently observed \cite{sturg1} that the photosynthetic
membranes in {\it Rsp. Photometricum} adapt to the light intensity
conditions under which they grow. Illuminated under High Light
Intensity (HLI) ($I_0\approx 100$W/m$^2$ where $I_0$ is the
growing light intensity), membranes grow with a ratio of
antenna-core complexes (i.e. stoichiometry) LH2/LH1$\approx$
3.5-4. For Low Light Intensity (LLI) ($I_0\approx 10$W/m$^2$),
this ratio increases to 7-9. The features that reveal an
unexpected change in the ratio of harvesting complexes, in
bacterias grown under HLI and LLI are shown in Fig.\ref{membs}(a)
and  Fig.\ref{membs}(b), respectively. Here we present a quantitative theory to explain this adaptation
 in terms of a dynamical interplay between excitation kinetics and
reaction-center dynamics. In particular, the paper lays out the model, its motivation and implications, in a progressive manner in order to facilitate understanding.  Although our model treats the excitation transport as a
noisy, classical process, we stress that the underlying quantities being
transported are quantum mechanical many-body excitations
\cite{fassioli,silbey}. The membrane architecture effectively acts
as a background network which loosely coordinates the entire
process.

\begin{figure}
\centering
\includegraphics[width=0.7\columnwidth]{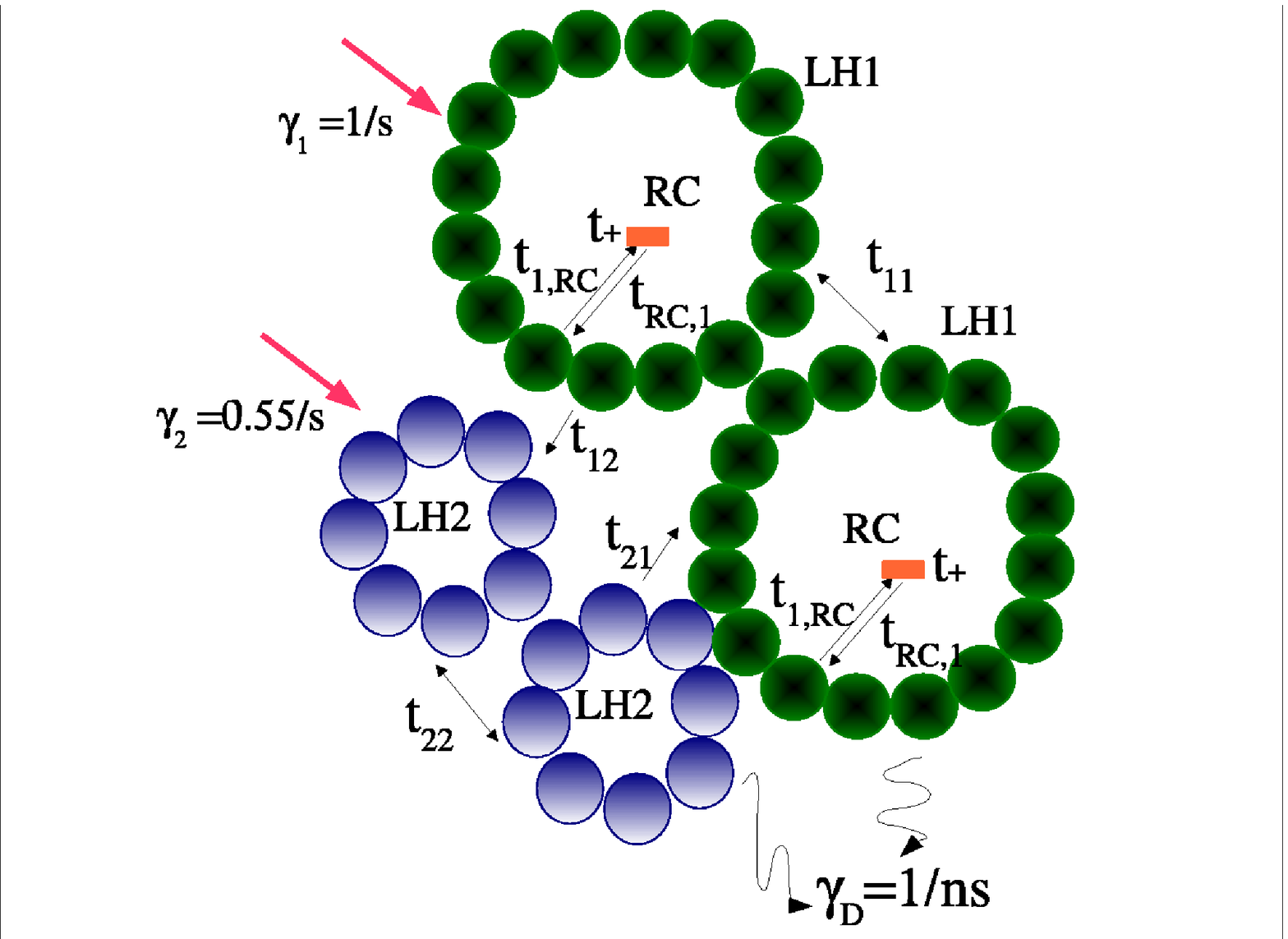}
\vspace{-0.01 cm} \caption{Schematic of the biomolecular
photosynthetic machinery in purple bacteria, together with
relevant inter-complex mean transfer times $t_{ij}$, dissipation
rate $\gamma_D$, and normalized light intensity rate
$\gamma_{1(2)}$}\label{rates}
\end{figure}

\section{Structure of complexes and excitation kinetics in small networks}\label{StructandKin}

Figure \ref{rates} summarizes the relevant biomolecular complexes
in purple bacteria {\it Rsp. Photometricum} \cite{sturgstoich},
together with timescales governing the excitation kinetics and
reaction center dynamics. Each LH2 can absorb light with
wavelengths 839 to 846 nm, while LH1 absorbs maximally at 883 nm.
The LH1 forms an ellipse which completely surrounds the reaction
center (RC) complex. Within the RC, a dimer of
bacterio-chlorophyls (BChls) known as the special pair P, can be
excited. The excitation (P$^*$) induces ionization  (P$^+$) of the
special pair, and hence metabolism. The initial photon absorption
is proportional to the complex cross-sections, which have been
calculated for LH1 and LH2 complexes \cite{francke95}.  With
$n(\lambda)$ incident photons of wavelength $\lambda$,  an 18
W/m$^2$ light intensity yields a photon absorption rate for
circular LH1 complexes in {\it Rb. sphaeroides} \cite{Geyer} given
by $\int n(\lambda) \sigma_{{\rm LH1}}(\lambda)d\lambda=18
\mbox{s}^{-1}$, where $\sigma_{{\rm LH1}}$ is the LH1 absorption
cross section.  For LH2 complexes, the corresponding photon
capture rate is $10 $s$^{-1}$. Extension to other intensity
regimes is straightforward, by normalizing to unit light
intensity. The rate of photon absorption normalized to 1 W/m$^2$
intensity, will be $\gamma_{1}=\frac{18}{18}=1$s$^{-1}$ for an
individual LH1, and $\gamma_2= \frac{10}{18}=0.55 $s$^{-1}$ for
individual LH2 complexes. The complete vesicle containing several
hundreds of complexes, will have an absorption rate $
\gamma_A=I(\gamma_1 N_1+\gamma_2N_2)$ where $ N_{1(2)}$ is the
number of LH1 (LH2) complexes in the vesicle, and $I$ is the light
intensity. The number of RC complexes is therefore also equal to $
N_{1}$. Excitation transfer occurs through induced dipole
transfer, among BChls singlet transitions.  The common
inter-complex BChl distances 20-100 $\AA$ \cite{sturg1,bahatyrova}
cause excitation transfer to arise through the Coulomb interaction
on the picosecond time-scale \cite{review}, while vibrational
dephasing destroys coherences within a few hundred femtoseconds
\cite{ecoh1,ecoh2}. The Coulomb interaction de-excites an
initially excited electron in the donor complex while
simultaneously exciting an electron in the acceptor complex. As
dephasing occurs, the donor and acceptor phase become
uncorrelated.  Transfer rate measures from pump-probe experiments
agree with generalized F\"orster calculated rates \cite{review},
assuming intra-complex delocalization. LH2$\rightarrow$LH2
transfer has not been measured experimentally, although an
estimate of $t_{22}=10$ ps has been calculated \cite{review}.
LH2$\rightarrow$ LH1 transfer has been measured for {\it R.
Sphaeroides} as $t_{21}= 3.3$ps \cite{Hess}. Due to formation of
excitonic states \cite{fassioli,silbey}, back-transfer
LH1$\rightarrow$ LH2 is enhanced as compared to the canonical
equilibrium rate for a two-level system, up to a value of
$t_{12}=15.5 $ps. The LH1$\rightarrow$LH1 mean transfer time
$t_{11}$ has not been measured, but generalized F\"orster
calculation \cite{ritz} has reported an estimated mean time
$t_{11}$ of 20 ps.  LH1$\rightarrow$ RC transfer occurs due to
ring symmetry breaking through second and third lowest exciton
lying states \cite{Damja}, as suggested by agreement with the
experimental transfer time of 35-37 ps  at 77 K
\cite{Bergstrom,Visscher}.  Increased spectral overlap at room
temperature improves the transfer time to $t_{1,RC}=25$ ps as
proposed by \cite{Vgrondelle}. A photo-protective design makes the
back-transfer from an RC's fully populated lowest exciton state to
higher-lying LH1 states occur in a calculated time of
$t_{RC,1}=$8.1 ps \cite{Damja}, close to the experimentally
measured 7-9 ps  estimated from decay kinetics after RC excitation
\cite{Timpmann}.  The first electron transfer step $P^*\rightarrow
P^+$ occurs in the RC within $t_{+}=$3 ps, used for quinol
($Q_BH_2$) production \cite{review}.  Fluorescence, inter-system
crossing, internal conversion and further dissipation mechanisms,
have been included within an effective single lifetime
$1/\gamma_D$ of 1 ns \cite{ritz}. Due to the small absorption
rates in $\gamma_A$, two excitations will only rarely occupy a
single harvesting structure -- hence it is sufficient to include
the ground $s=0$ and one exciton states $s=1$ for each harvesting
complex.

We now introduce the theoretical framework that we use to describe
the excitation transfer, built around the experimental and
theoretical parameters just outlined.  In the first part of the
paper, our calculations are all numerical -- however we turn to an
analytic treatment in the latter part of the paper. We start by
considering a collective state with $N=N_2+2N_1$ sites --
resulting from $N_1$ LH1s, $N_2$ LH2s and hence $N_1$ RC complexes
in the vesicle --  in terms of a set of states having the form
$\{s_1,..,s_N\}$ in which any complex can be excited or unexcited,
and a maximum of $N$ excitations can exist in the membrane. If
only excitation kinetics are of interest, and only two states
(i.e. excited and unexcited) per complex are assumed, the set of
possible states has $2^N$ elements. We introduce a vector
$\vec{\rho}=(\rho_1,\rho_2,...,\rho_{2^N})$ in which each element
describes the probability of occupation  of a collective state
comprising several excitations. Its time evolution obeys a master
equation
\begin{equation}\label{meq}
\partial_t\rho_i(t)=\sum_{j=1}^{2^N}G_{i,j}\rho_j(t).
\end{equation}
Here $G_{i,j}$ is the transition rate from a site $i$ to a site
$j$.  Since the transfer rates do not depend on time, this yields
a formal solution $\vec{\rho}(t)=\tilde{e^{G t}}\vec{\rho}(0)$.
Small absorption rates lead to single excitation dynamics in the
whole membrane, reducing the size of $\vec{\rho}(t)$ to the total
number of sites $N$. The probability to have one excitation at a
given complex initially, is proportional to its absorption cross
section, and can be written as
$\vec{\rho}(0)=\frac{1}{\gamma_A}(\underbrace{\gamma_1,...}_{N_1},\underbrace{\gamma_2,...}_{N_2},\underbrace{0,..}_{N_1})$,
where subsets correspond to the $N_1$ LH1s, the $N_2$ LH2s and the
$N_1$ RCs respectively.  Our interest lies in $\hat{p}_k$ which
is the normalized probability to find an excitation at a complex,
given that at least one excitation resides in the network:
\begin{equation}
\hat{p}_{k}(t)=\frac{\rho_{k}(t)}{\sum_{i=1}^{N} \rho_{i}(t)}\ \ .
\end{equation}

\begin{figure}
\centering
\includegraphics[width=0.3\columnwidth]{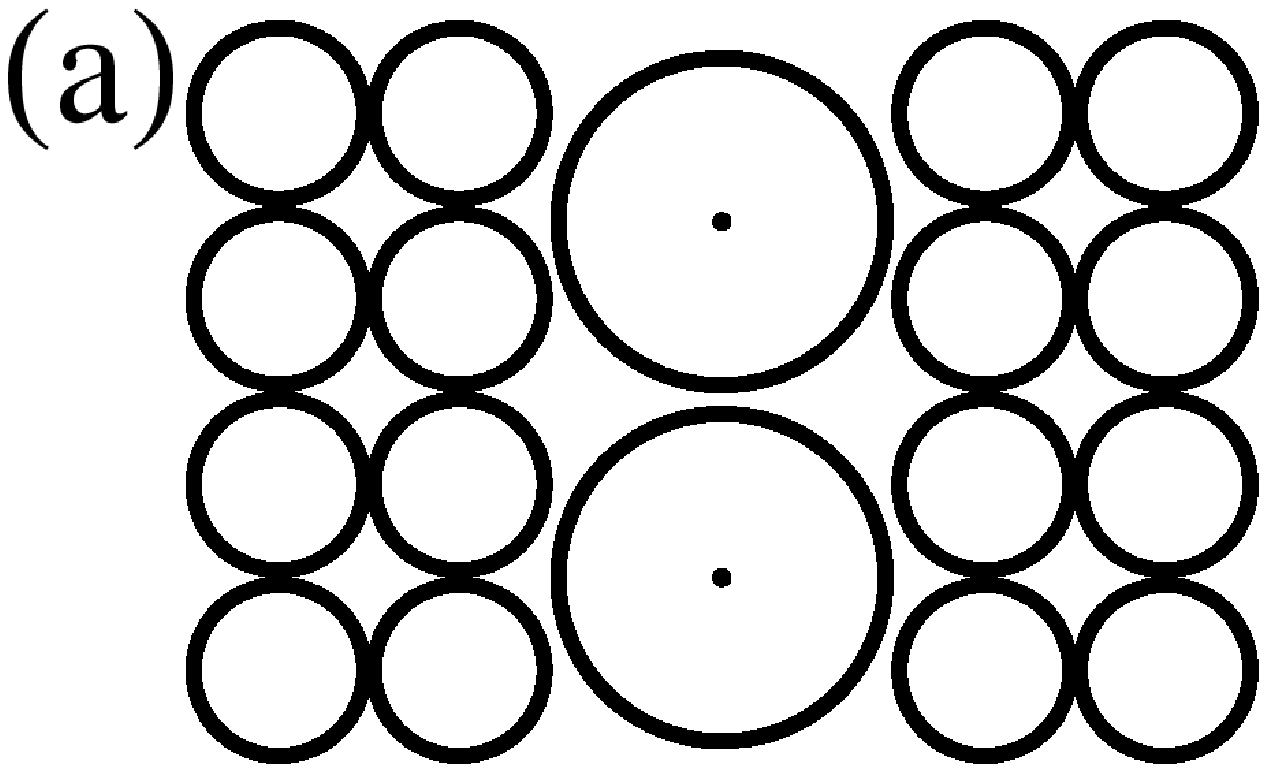}
\includegraphics[width=0.3\columnwidth]{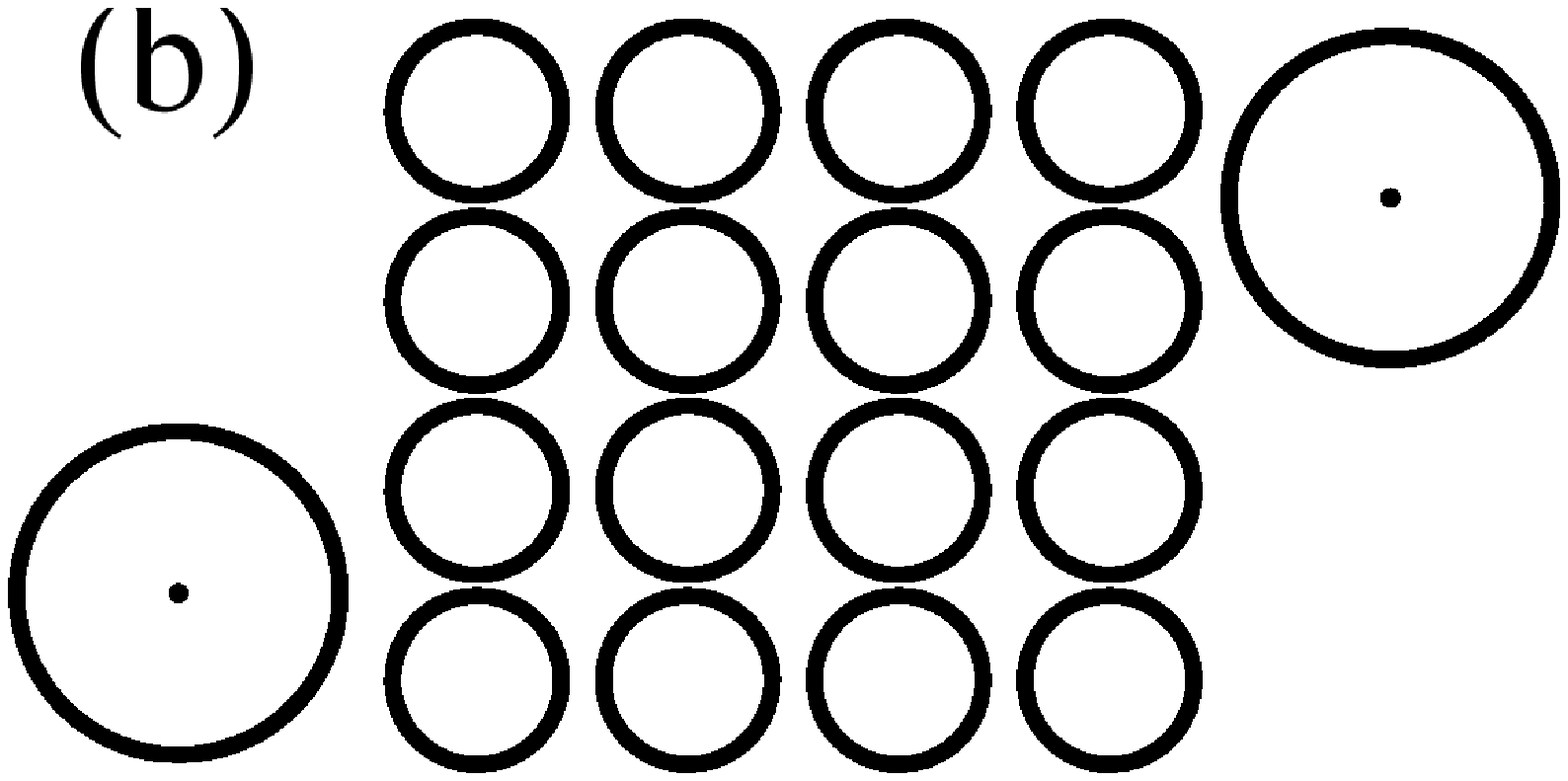}
\includegraphics[width=0.3\columnwidth]{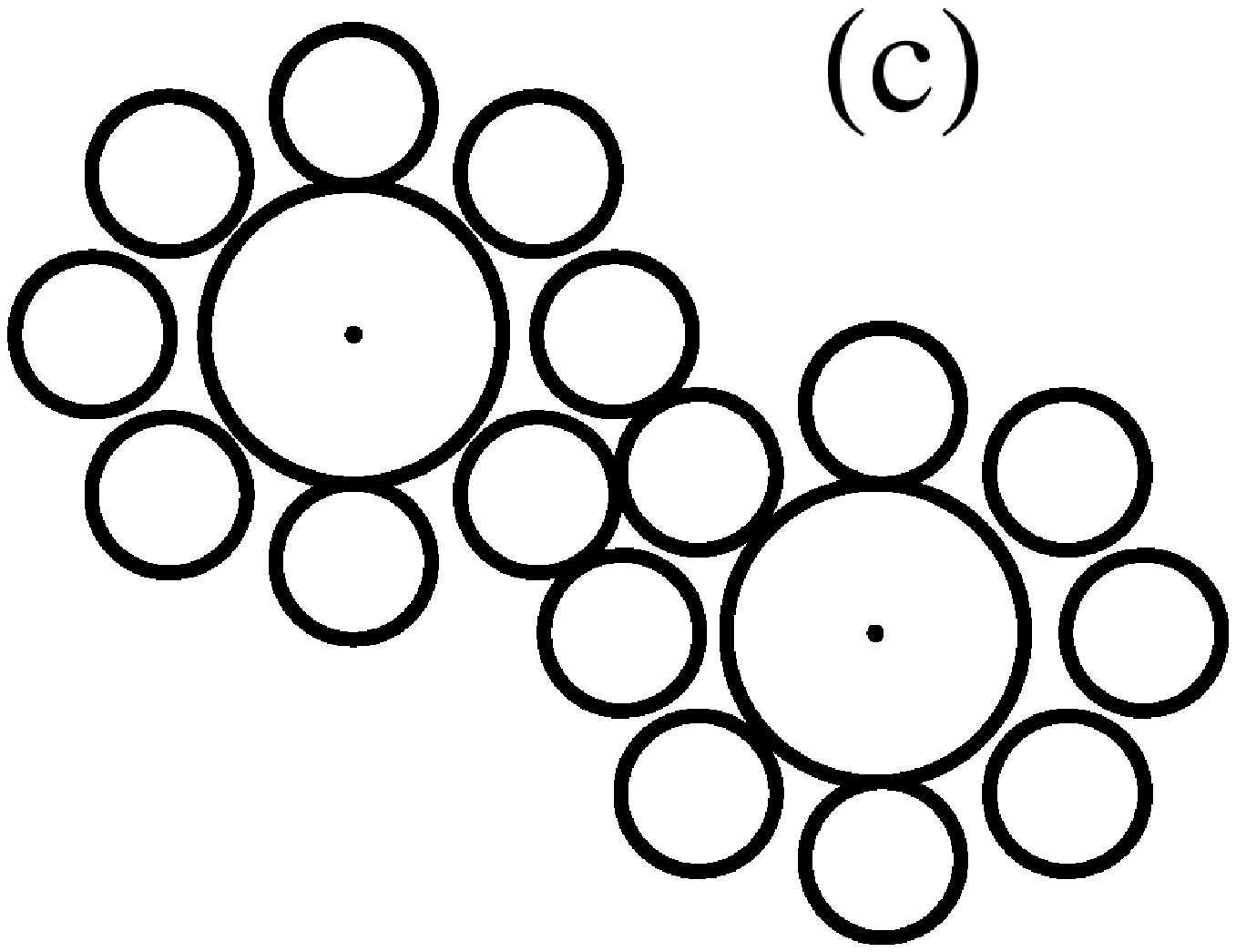}\\
\begin{minipage}{0.65\columnwidth}
\hspace{-2 cm}
\includegraphics[width=1\columnwidth]{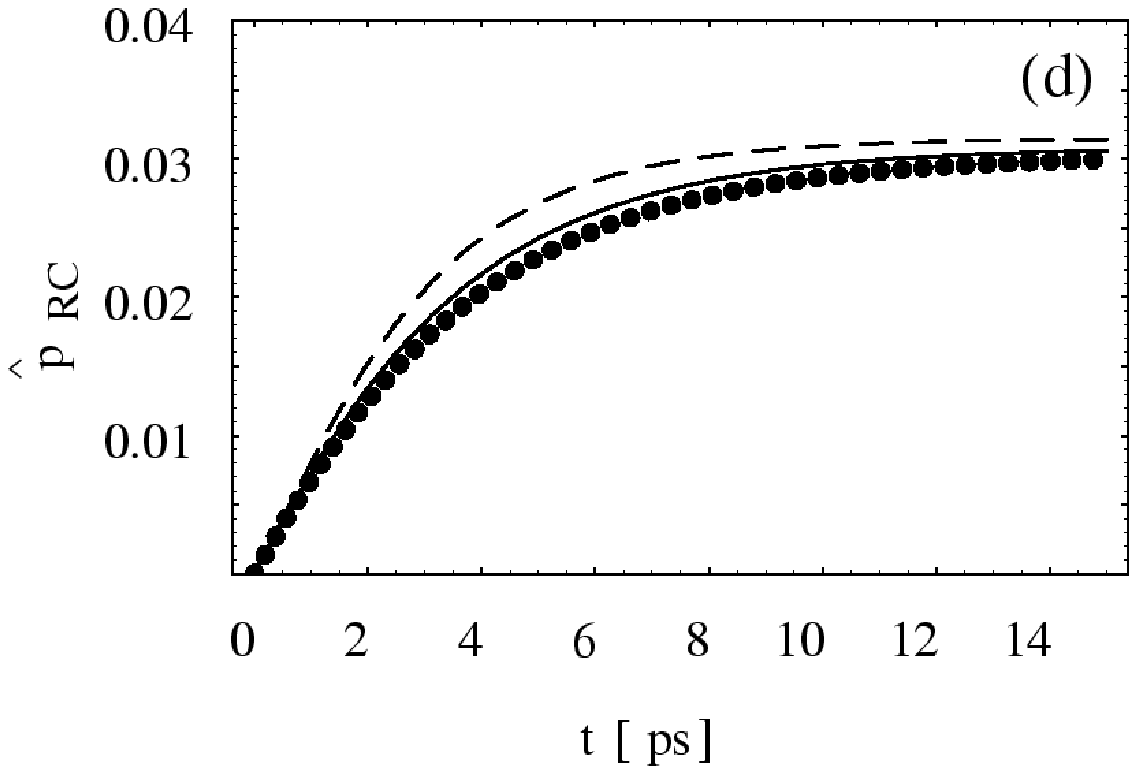}
\end{minipage}\hspace{-2 cm}
\begin{minipage}{0.31\columnwidth}
\includegraphics[width=1\columnwidth]{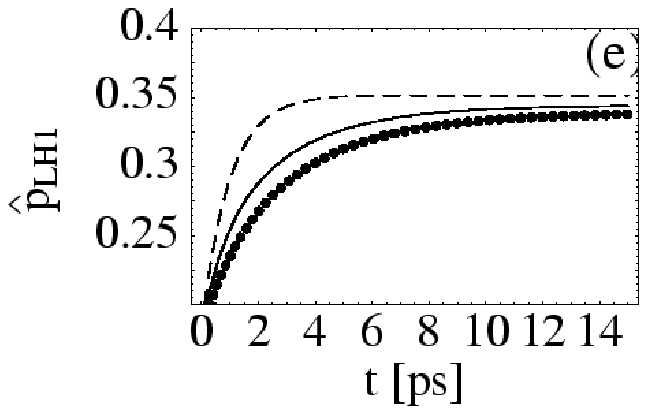}
\includegraphics[width=1\columnwidth]{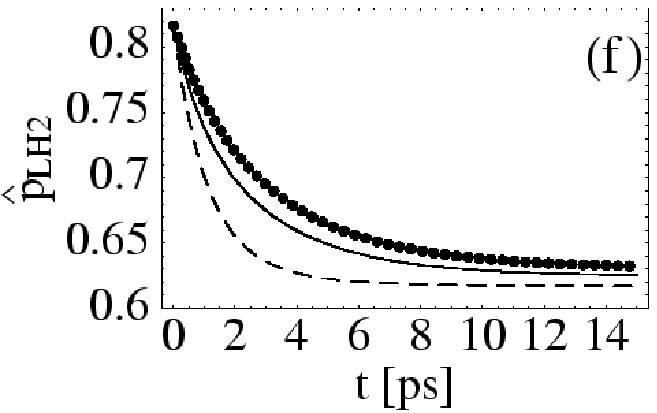}
\end{minipage}
\caption{Top panel: Three example small network architectures. The bottom panel shows the normalized probabilities for finding an
excitation at an RC (see (d)), an LH1 (see (e)), or an LH2 (see (f)). In panels (d)-(f), we represent
these architectures as follows: (a) is a continuous line; (b) is a dotted line; (c) is a dashed line.}\label{archs0}
\end{figure}

In order to appreciate the effects that network architecture might
have on the model's dynamics, we start our analysis by studying
different arrangements of complexes in small model networks,
focusing on architectures which have the same amount of LH1, LH2
and RCs as shown in the top panel of Fig.\ref{archs0}(a), (b) and
(c). The bottom panel Fig.\ref{archs0} (d)-(f) shows that
$\hat{p}_{k}$ values for RC, LH1 and LH2 complexes, respectively.
Fig.\ref{archs0}(d) shows that the highest RC population is
obtained in configuration (c), followed by configuration (a)  and
(b) whose ordering relies in the connectedness of LH1s to antenna
complexes. Clustering of LH1s will limit the number of links to
LH2 complexes, and reduce the probability of RC ionization. For
completeness, the probability of occupation in LH1 and LH2
complexes  (Figs.\ref{archs0}(e) and (f), respectively), shows
that increased RC occupation benefits from population imbalance
between LH1 enhancement and LH2 reduction. As connections among
antenna complexes become more favored, the probability of finding
an excitation on antenna complexes will become smaller, while the
probability of finding excitations in RCs is enhanced.

\begin{figure}
\centering
\includegraphics[width=0.45\columnwidth]{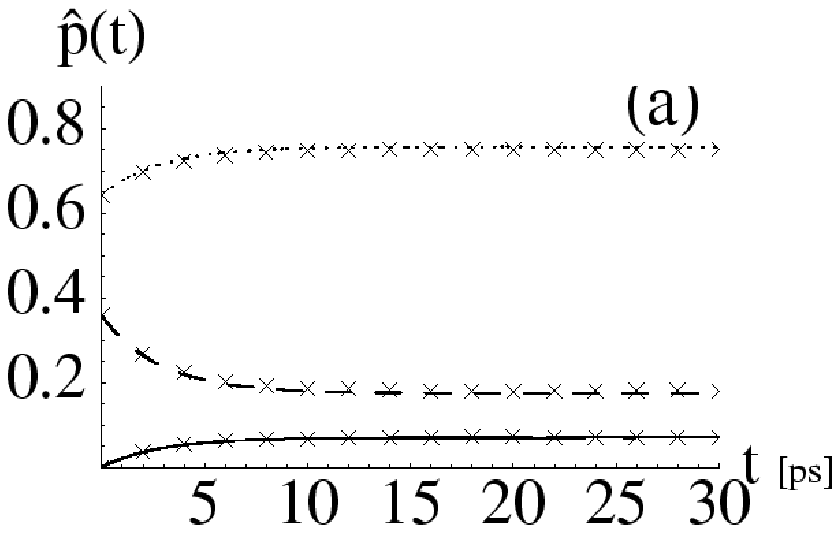}
\includegraphics[width=0.45\columnwidth]{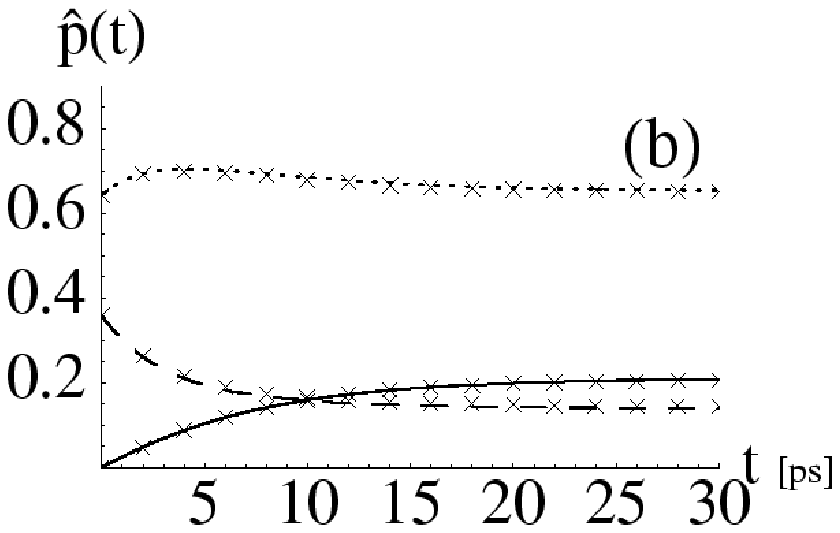}
\caption{Normalized probabilities $\hat{p}_{k}$ for finding the
excitation at an LH2 (dashed), LH1 (dotted)  or at an RC
(continuous), for (a) $t_{+}=3$ps, and (b) $t_{+}\rightarrow
\infty$. Crosses are the results from the Monte Carlo
simulation.}\label{green}
\end{figure}

This discussion of simple network architectures, provides us with
a simple platform for testing the notion of energy funneling,
which is a phenomenon that is commonly claimed to arise in such
photosynthetic structures. We start with a minimal configuration
corresponding to a basic photosynthetic unit: one LH2, one LH1 and
its RC.  Figure \ref{green}(a) shows that excitations will mostly
be found in the LH1 complex, followed by occurrences at the LH2
and lastly at the RC. Figure \ref{green}(b) shows clearly the
different excitation kinetics which arise when the RC is initially
unable to start the electron transfer $P^*\rightarrow P^+$, and
then after $\approx 15$ps the RC population increases with respect
to the LH2's. This confirms that the energy funneling concept is
valid for these small networks \cite{review,ritz}, i.e.
excitations have a preference to visit the RC ($t_{1,RC}=25$ps) as
compared to being transferred to the light-harvesting complexes
($t_{12}=15.5$ps). However, in natural scenarios involving entire
chromatophores with many complexes, we will show that energy
funneling is not as important due to increased number of available
states, provided from all LH2s surrounding a core complex.

Given the large state-space associated with such multiple
complexes, our subsequent model analysis will be based on a
discrete-time random walk for  excitation hopping between
neighboring complexes. In particular, we use a Monte Carlo method
to simulate the events of  excitation  transfer, the photon
absorption and dissipation,  and the RC electron transfer. We have
checked that our Monte Carlo simulations accurately reproduce the
results of the population-based calculations described above, as
can be seen from Figs.\ref{green}(a) and (b).

\section{Performance measures of complete chromatophore vesicles} \label{completelocal}

We now turn to discuss the application of the model to the
empirical biological structures of interest, built from the three
types of complex $k$ (LH1, $k$=1; LH2, $k$=2; RC, k=3). In
particular, we have carried out extensive simulations to
investigate the role of the following quantities in the complete
chromatophore vesicles:
\begin{itemize}
\item{Adjacency geometry of LH1s and LH2s. The LH2s  are more abundant than LH1s and both complexes tend to
form clusters, while LH2s are also generally found surrounding the
LH1s.} \item {The average time an excitation spends $\hat{t}_k$ in
complex type $k$.} \item{The probability $p_{R_k}$ of finding an
excitation in complex type $k$.} \item{Dissipation $d_i$ which measures
the probability for  excitations to dissipate at site $i$, from
which  the probability $D_{k}$ of dissipation in core or antenna
complexes can be obtained by adding all $d_i$ concerning complex type
$k$.} \item{The sum over all complexes of the dissipation probability, which
gives the probability for an excitation to be dissipated. The
efficiency of the membrane is the probability of using an excitation
in any RC, i.e. $\eta=1-\sum_i d_i $.}
\end{itemize}

\begin{figure}
\includegraphics[width=0.45\columnwidth]{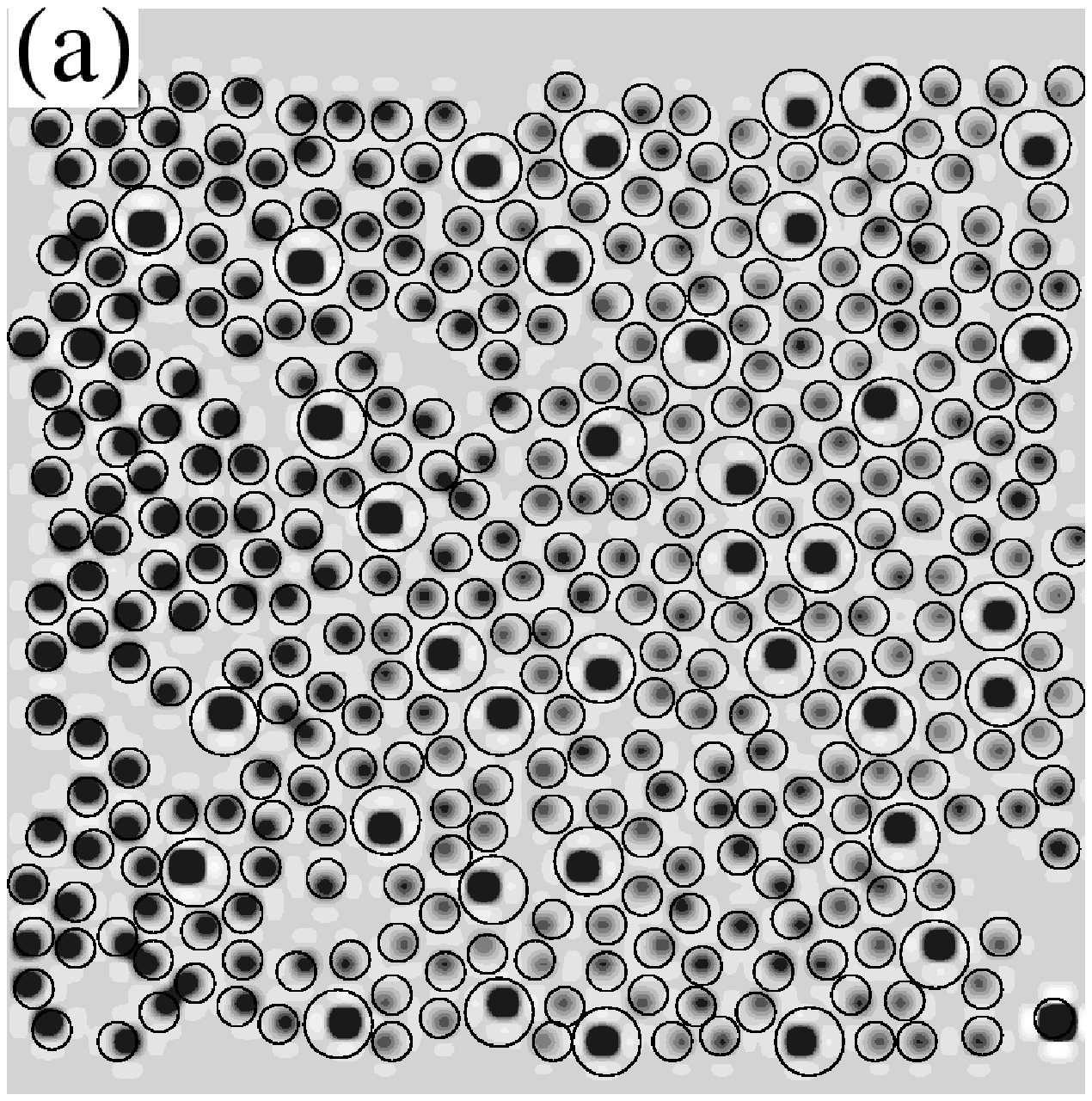}
\includegraphics[width=0.45\columnwidth]{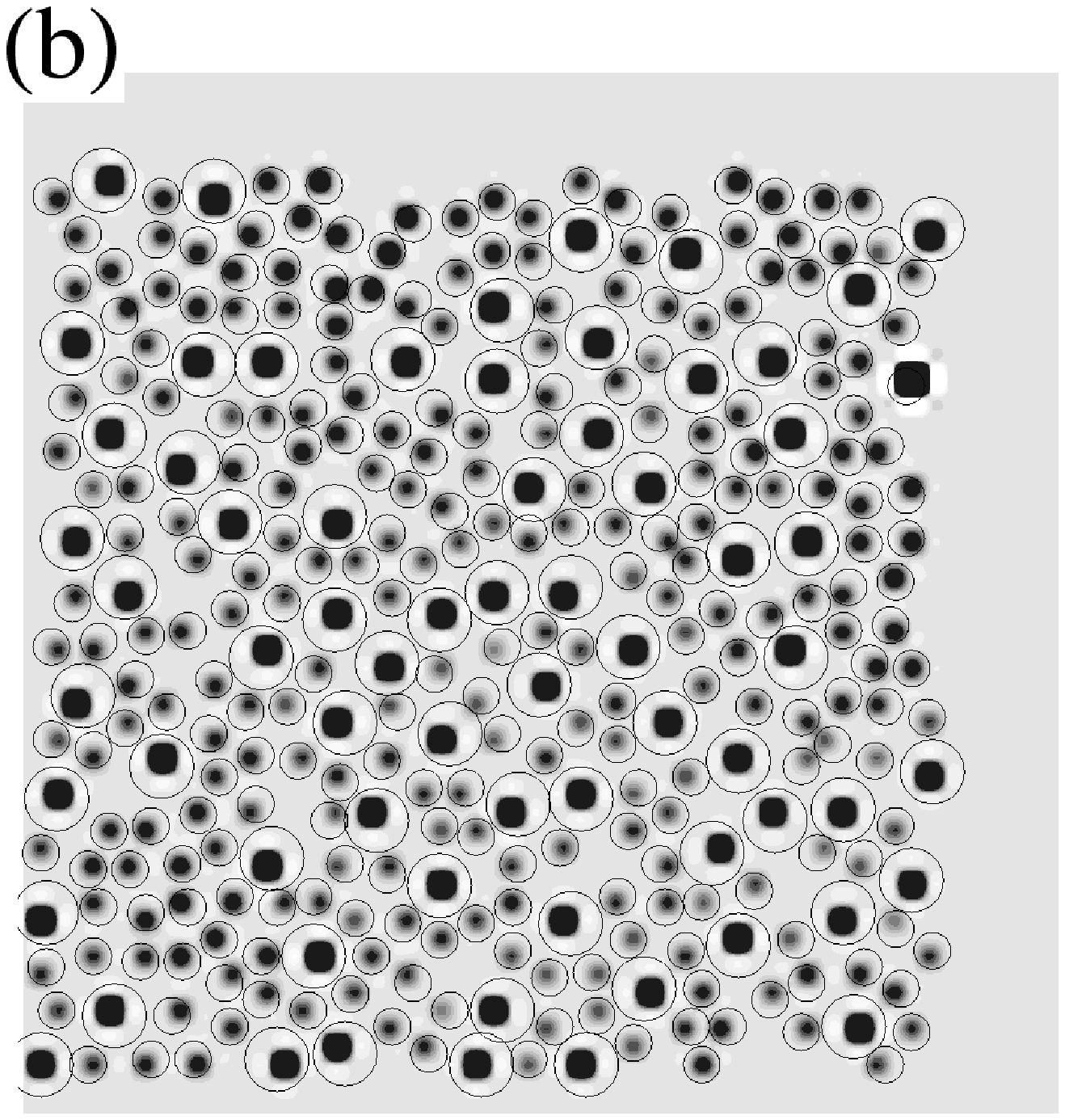}
\caption{Contour plots for dissipation in  LLI (a) and HLI (b)
membranes. Greater contrast means higher values for dissipation.
The simulation is shown after $10^6$ excitations were absorbed by
the membrane with rate $\gamma_A$.}\label{dissmembranes}
\end{figure}

Figure \ref{dissmembranes}(a) shows that the membrane grown under low light intensity (LLI) has highly
dissipative clusters of LH2s, in contrast to the uniform dissipation
in the high light intensity (HLI)  membrane (see Fig.\ref{dissmembranes}(b)).  This result is
supported by a tendency for excitations to reside longer in  LH2
complexes far from core centers (not shown), justifying the view
of LH2 clusters as excitation reservoirs.  However, for LLI and
HLI, the dissipation in LH1 complexes is undistinguishable.
In Table \ref{table1} we show the observables obtained using our numerical simulations. These show that:

\begin{enumerate}
\item{Funneling of excitations:}
\begin{itemize}
\item{The widely held view of the funneling of excitations to LH1
complexes, turns out to be a small network effect, which by no means reflects the
behavior over the complete chromatophore. Instead, we find
that excitations are found residing mostly in LH2 complexes.}
\item{Since a few  LH2s surround each LH1, the mean residence
times $t_{k}$ in all complexes is very similar.}
\end{itemize}
\item{Dissipation and performance:}
\begin{itemize}
\item{Excitations are dissipated more efficiently in individual
LH1 complexes, since  $\frac{D_{1}}{N_1}> \frac{D_{2}}{N_2}$.}
\item{Dissipation in a given complex type depends primarily
on its relative abundance, since  $\frac{D_{k}}{D_j}\approx
\frac{N_k}{N_j}$} \item{HLI membranes are more efficient than LLI
membranes.}
\end{itemize}
\end{enumerate}

\begin{table}[hh]
\begin{center}
\begin{tabular}{|r|rrrrrrrrrrr|}
\hline
Membrane & $\hat{t}_{2}$& $\hat{t}_{1}$&$p_{R_2}$ & $p_{R_1}$& $D_{2}$ & $D_{1}$ &  $\frac{D_{2}}{N_2}$ &  $\frac{D_{1}}{N_1}$ & $\frac{D_{2}}{D_{1}}$ & $s=\frac{N_2}{N_1}$&$\eta=\frac{n_{RC}}{n_A}$\\
\hline
LLI & 2.22& 2.39&0.72& 0.25 &0.74 & 0.26 &  2.2 & 7.2 & 9.13 & 9.13 &0.86\\
HLI &1.70& 2.65& 0.50 & 0.46 & 0.52 & 0.48& 1.9 & 7.1 & 3.88 & 3.92 &0.91\\
\hline
\end{tabular}
\end{center}
\caption{Residence time $\hat{t}_{k}$ (in picoseconds),
dissipation $D_k$, residence probability $p_{R_k}$, unitary
dissipation per complex  $\frac{D_{k}}{N_k}$ ($\times 10^{-3}$),
on $k=\{1,2\}$ corresponding to $N_1$ LH1 and $N_2$ LH2 complexes
respectively. Stoichiometry $s$ and efficiency $\eta$ are also
shown.}\label{table1}
\end{table}

\begin{figure}
\includegraphics[width= 7 cm]{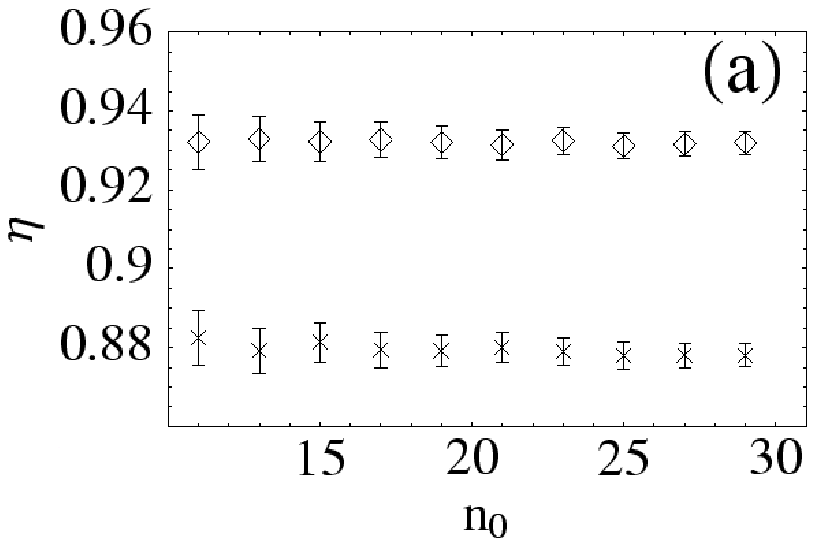}
\includegraphics[width=7 cm]{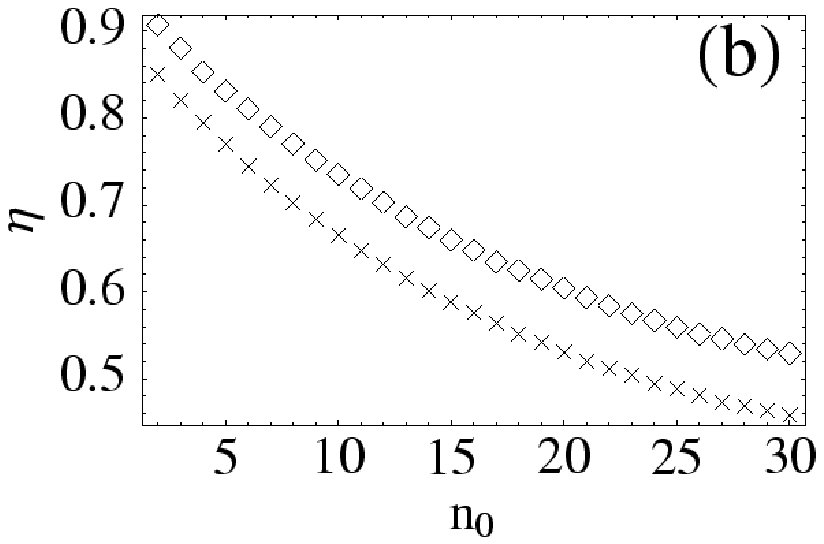}
\caption{Efficiency of multiple excitation dynamics: (a) blockade
and (b) annihilation mechanisms, for LLI (crosses) and HLI
(diamonds) membranes. $n_0$ corresponds to the initial number of
excitations in each realization.}\label{multiple}
\end{figure}

For the present discussion, the most important finding from our
simulations is that the adaptation of purple bacteria does {\em
not} lie in the single excitation kinetics. In particular, LLI
membranes are seen to reduce their efficiency globally at the
point where photons are becoming scarcer -- hence the answer to
adaptation must lie in some more fundamental trade-off (as we will
later show explicitly). Due to the dissimilar timescales between
millisecond absorption \cite{Geyer} and nanosecond dissipation
\cite{review},  multiple excitation dynamics are also unlikely to
occur within a membrane. However we note that simulations
involving  multiple excitations,  that include   blockade
(Fig.\ref{multiple}(a)) in which two excitations can not occupy
the same site,  does not appreciably lower the  efficiency $\eta$
up to thirty excitations. We find that annihilation
(Fig.\ref{multiple}(b)), in which two excitations annihilate when
they occupy the site at the same time, diminishes the membrane's
performance equally in both HLI and LLI membranes.

\begin{figure}
\centering
\includegraphics[width=0.5\columnwidth]{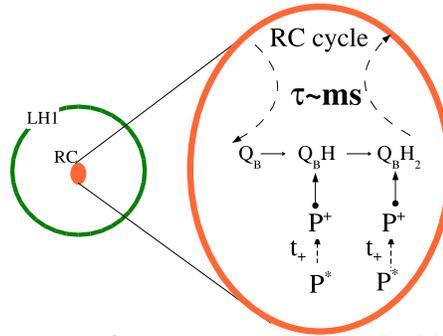}
\vspace{-1.7 cm} \caption{Reaction Center cycle, showing double
reduction of the special pair $P$ together with formation of
quinol $Q_BH_2$. There is a dead time $\tau$ on the millisecond
time-scale, before a new quinone $Q_B$ becomes
available.}\label{rccycle}
\end{figure}

\begin{figure}
\centering
\includegraphics[width=0.65\columnwidth]{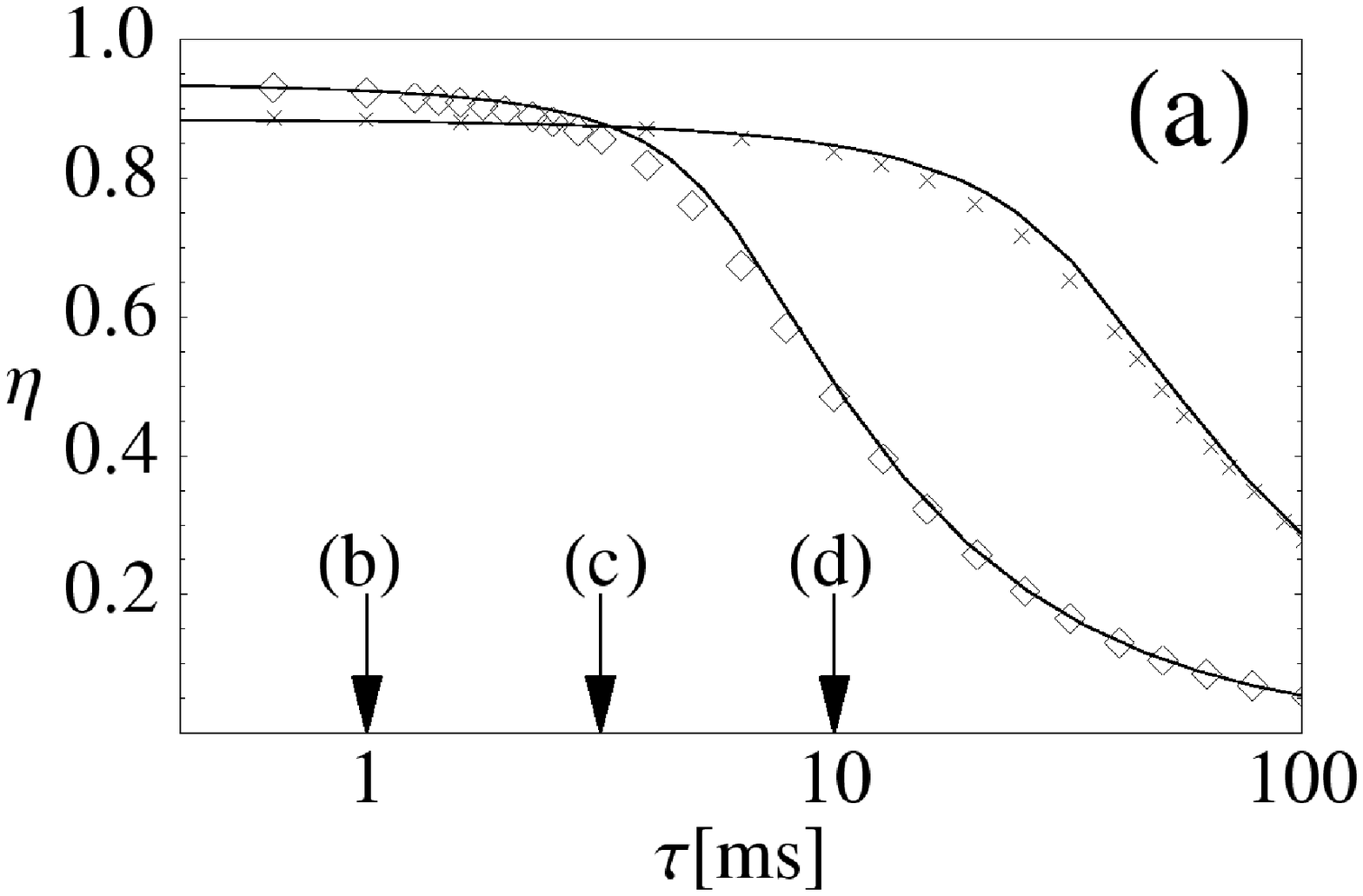}\\
\includegraphics[width=0.28\columnwidth]{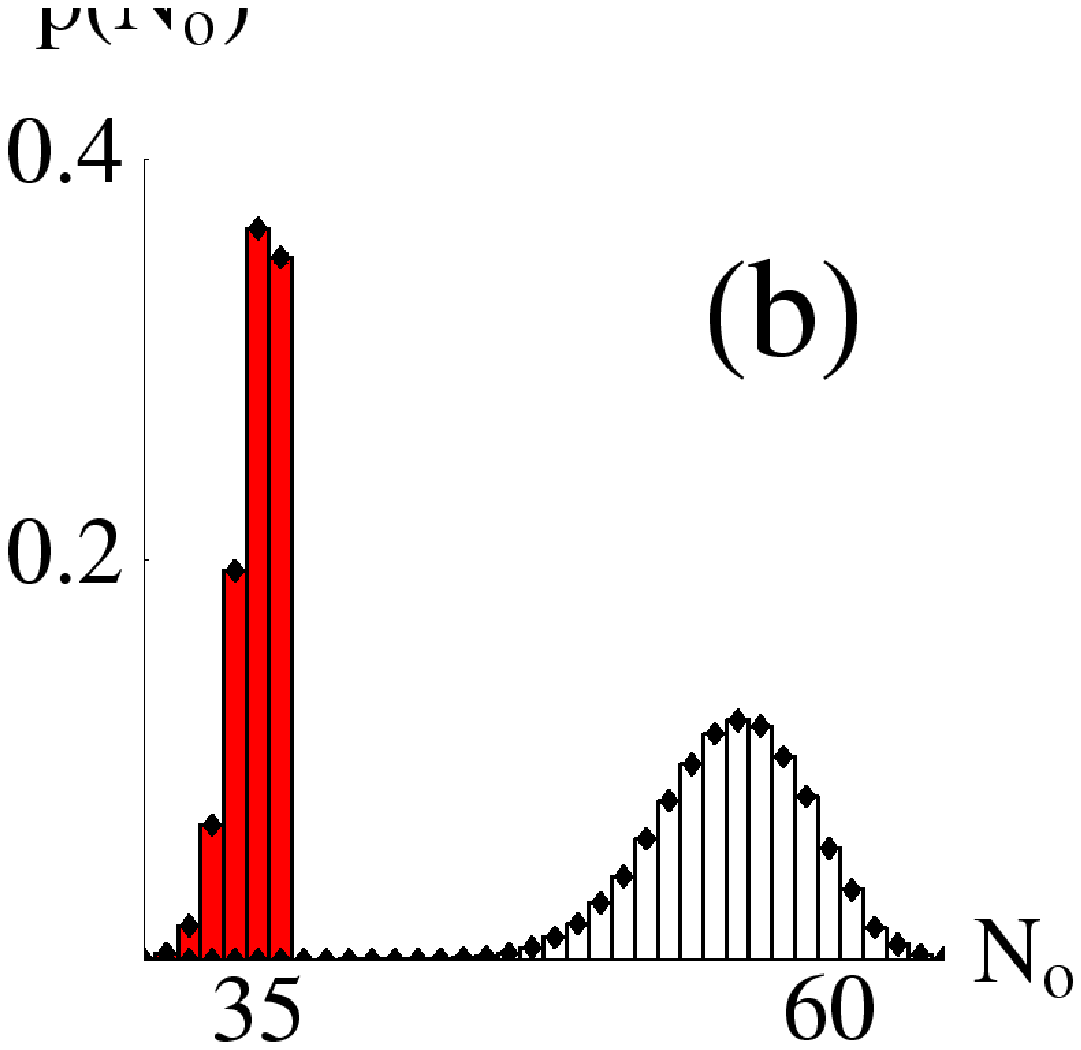}
\hspace{-0.4cm}
\includegraphics[width=0.28\columnwidth]{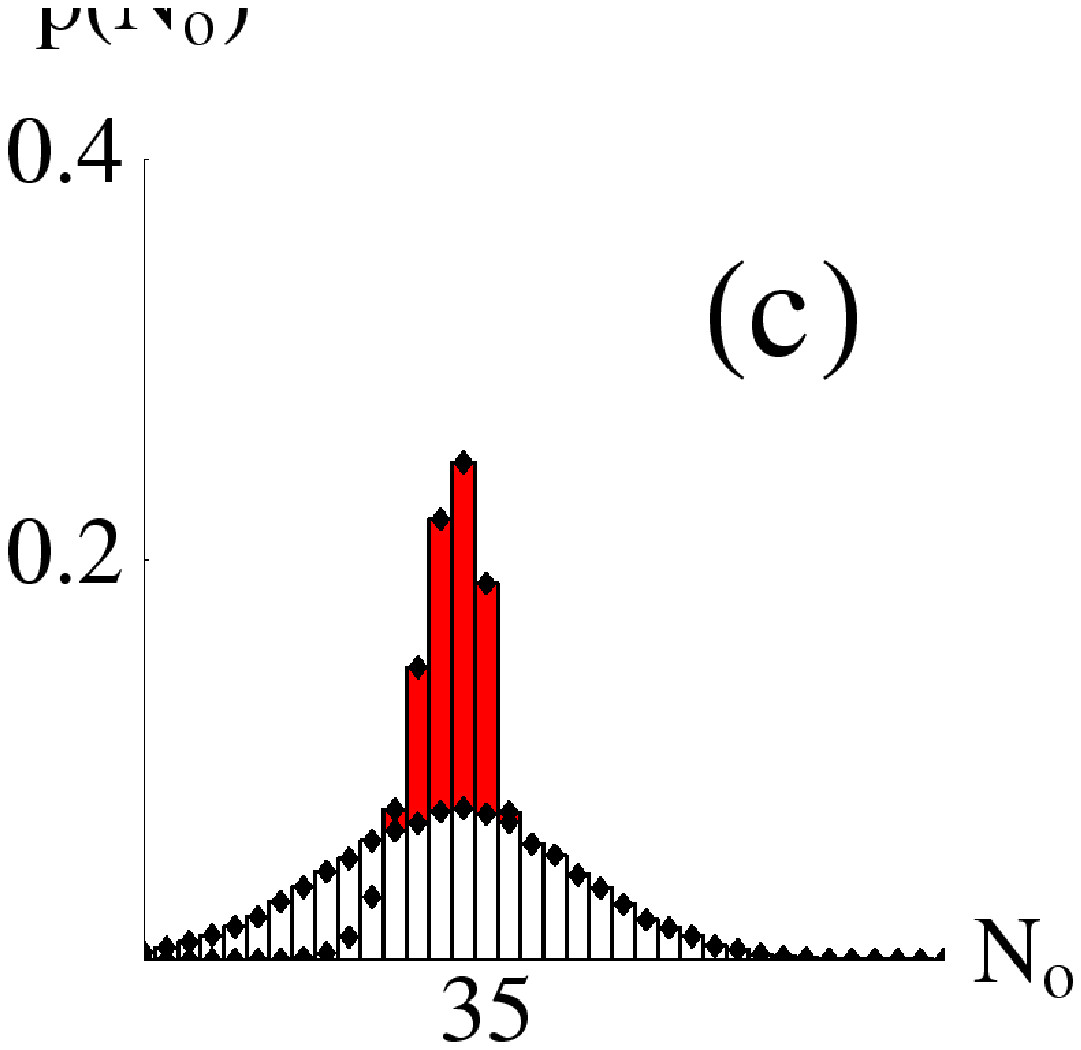}\hspace{-0.4cm}
\includegraphics[width=0.28\columnwidth]{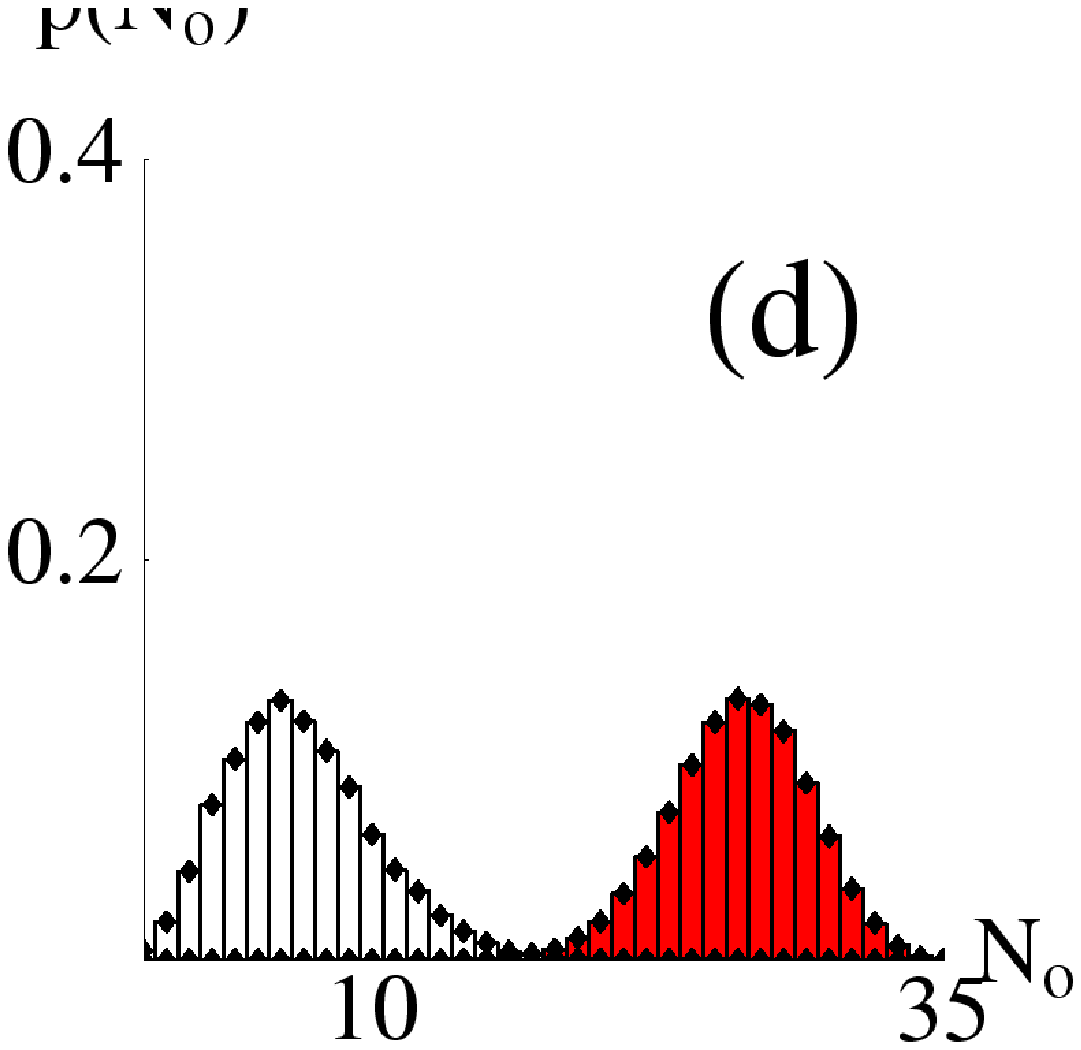}\\
\vspace{-.3cm} \caption {(a) Monte  Carlo  calculation of
efficiency $\eta$ of HLI (diamonds) and LLI (crosses) grown
membranes, as a function of the RC-cycling time $\tau$. Continuous
lines give the result of the analytical model.  (b), (c) and (d)
show  the distributions $p(N_o)$  of the number of open RCs for
the times shown with arrows in the main plot for HLI (filled bars)
and LLI (white bars).}\label{etatau}
\end{figure}

Our findings above show that the explanation for the observed
architecture adaptations (HLI and LLI) neither  lies in the
frequently quoted side-effect of multiple excitations, nor in the
excitation dynamics alone. Instead, as we now explain, the answer
as to how adaptation can prefer the empirically observed HLI and
LLI structures under different illumination conditions, lies in
the {\em interplay} between the excitation kinetics and
reaction-center cycling dynamics. By virtue of quinones-quinol and
cytochrome charge carriers, the RC dynamics features a `dead' (or
equivalently `busy') time interval during which quinol is
produced, removed and then a new quinone becomes available
\cite{RC1,RC2}. A single oxidation $P^*\rightarrow P^+$ will
produce $Q_BH$ in the reaction $Q_B\rightarrow Q_B^-\rightarrow
Q_BH$, and a second oxidation will produce quinol $Q_BH_2$ in the
reaction  $Q_BH\rightarrow Q_BH^-\rightarrow Q_BH_2$. Once quinol
is produced, it leaves the RC and a new quinone becomes attached.
The cycle is depicted in Fig. \ref{rccycle}, and is described in
the simulation algorithm by closing an RC for a time $\tau$ after
two excitations form quinol. This RC cycling time $\tau$ implies
that at any given time, not all RCs are available for turning the
electronic excitation into a useful charge separation. Therefore,
the number of useful RCs decreases with increasing $\tau$. Too
many excitations will rapidly close RCs, implying that any
subsequently available nearby excitation will tend to wander along
the membrane and eventually be dissipated - hence reducing $\eta$.
For the configurations resembling the empirical architectures
(Fig.\ref{membs}), this effect is shown  as a function of $\tau$
in Fig. \ref{etatau}(a)  yielding a wide range of RC-cycling times
at which LLI membrane is more efficient than HLI. Interestingly,
this range corresponds to the measured time-scale for $\tau$ of
milliseconds \cite{RC1,RC2}, and supports the suggestion that
bacteria improve their performance in LLI conditions by enhancing
quinone-quinol charge carrier dynamics as opposed to manipulating
exciton transfer. A recent proposal \cite{sturgdiff} has shown
numerically that the formation of LH2 para-crystalline domains
produces a clustering trend of LH1 complexes with enhanced quinone
availability -- a fact that would reduce the RC cycling time.
However, the  crossover of efficiency at $\tau\approx 3$ ms
implies that even if no enhanced RC-cycling occurs, the HLI will
be less efficient than the LLI membranes on the observed $\tau$
time-scale. The explanation is quantitatively related to the
number $N_o$ of open RCs.  Figs. \ref{etatau}(b), (c) and (d)
present the distribution $p(N_o)$ of open RCs, for both HLI and
LLI membranes and for the times shown with arrows in
Fig.\ref{etatau}(a).  When  the RC-cycling is of no importance
(Fig. \ref{etatau}(b)) almost all RCs remain open, thereby making
the HLI membrane more efficient than LLI  since having more (open)
RCs induces a higher probability for special pair oxidation. Near
the crossover in Fig. \ref{etatau}, both membranes have
distributions $p(N_o)$ centered around the same value (Fig.
\ref{etatau}(c)), indicating that although more RCs are present in
HLI vesicles, they are more frequently closed due to the ten fold
light intensity difference, as compared to LLI conditions. Higher
values of $\tau$ (Fig. \ref{etatau}(d)) present  distributions
where the LLI has more open RCs, in order to yield a better
performance when photons are scarcer. Note that distributions
become wider when RC cycling is increased, reflecting the
mean-variance correspondance of Poissonian statistics used for
simulation of $\tau$.  Therefore the trade-off between RC-cycling,
the actual number of RCs and the light intensity, determines the
number of open RCs and hence the performance of a given
photosynthetic vesicle architecture (i.e. HLI versus LLI).  Guided
by the Monte Carlo  numerical results, we develop in Sec.
\ref{analyt} an analytical model (continuous lines in
Fig.\ref{etatau}) that supports this discussion.

\begin{figure}
\centering
\includegraphics[width=0.65\columnwidth]{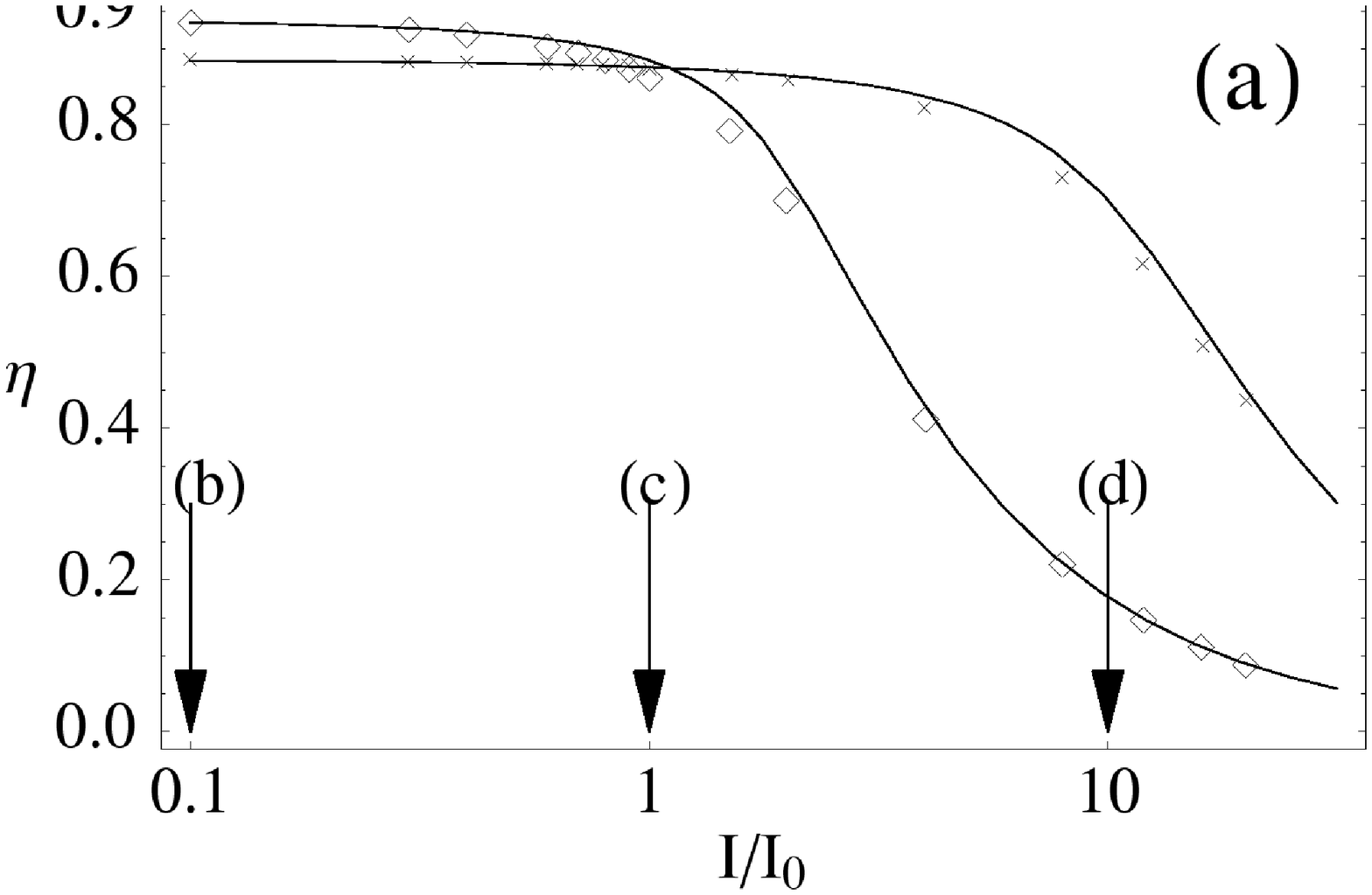}\\
\includegraphics[width=0.28\columnwidth]{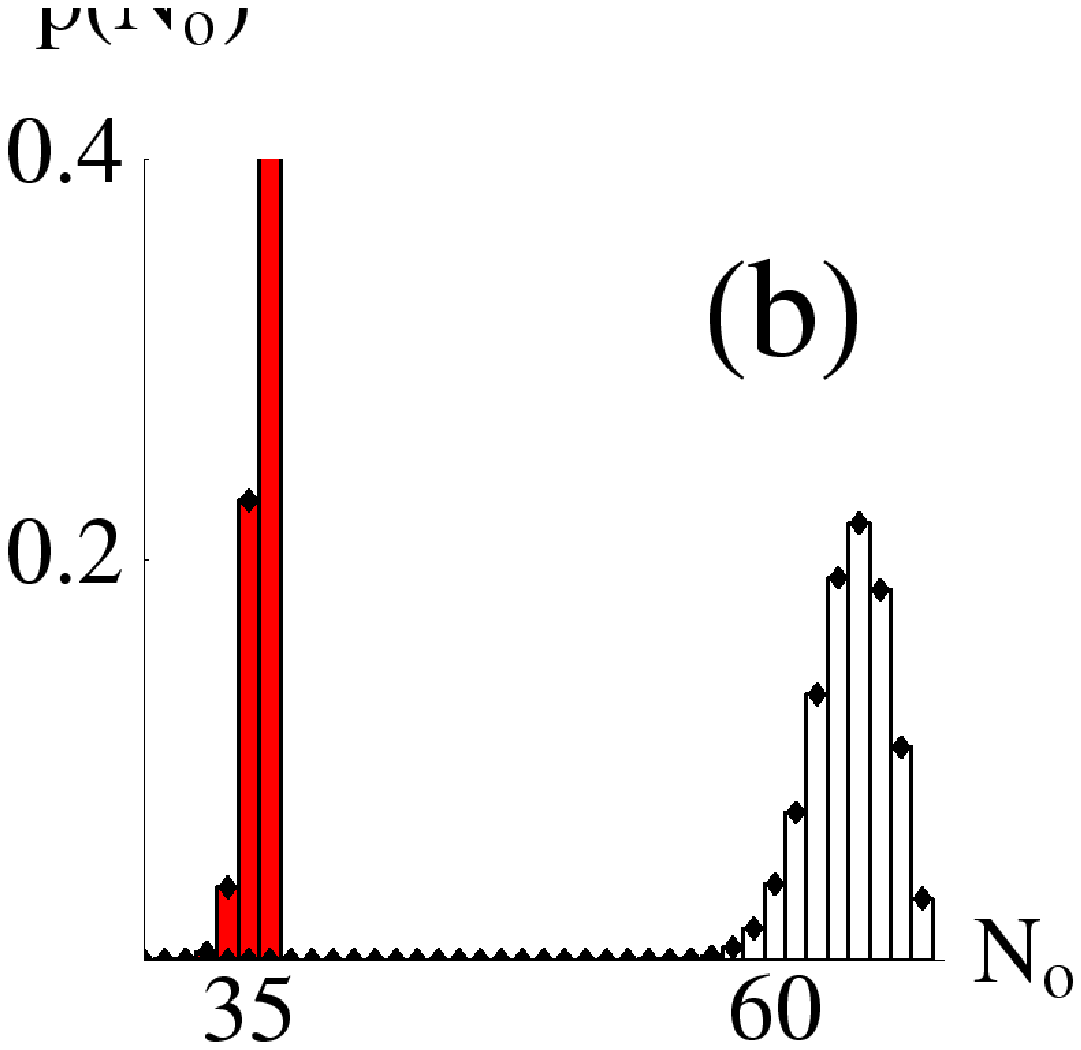}
\hspace{-0.4cm}
\includegraphics[width=0.28\columnwidth]{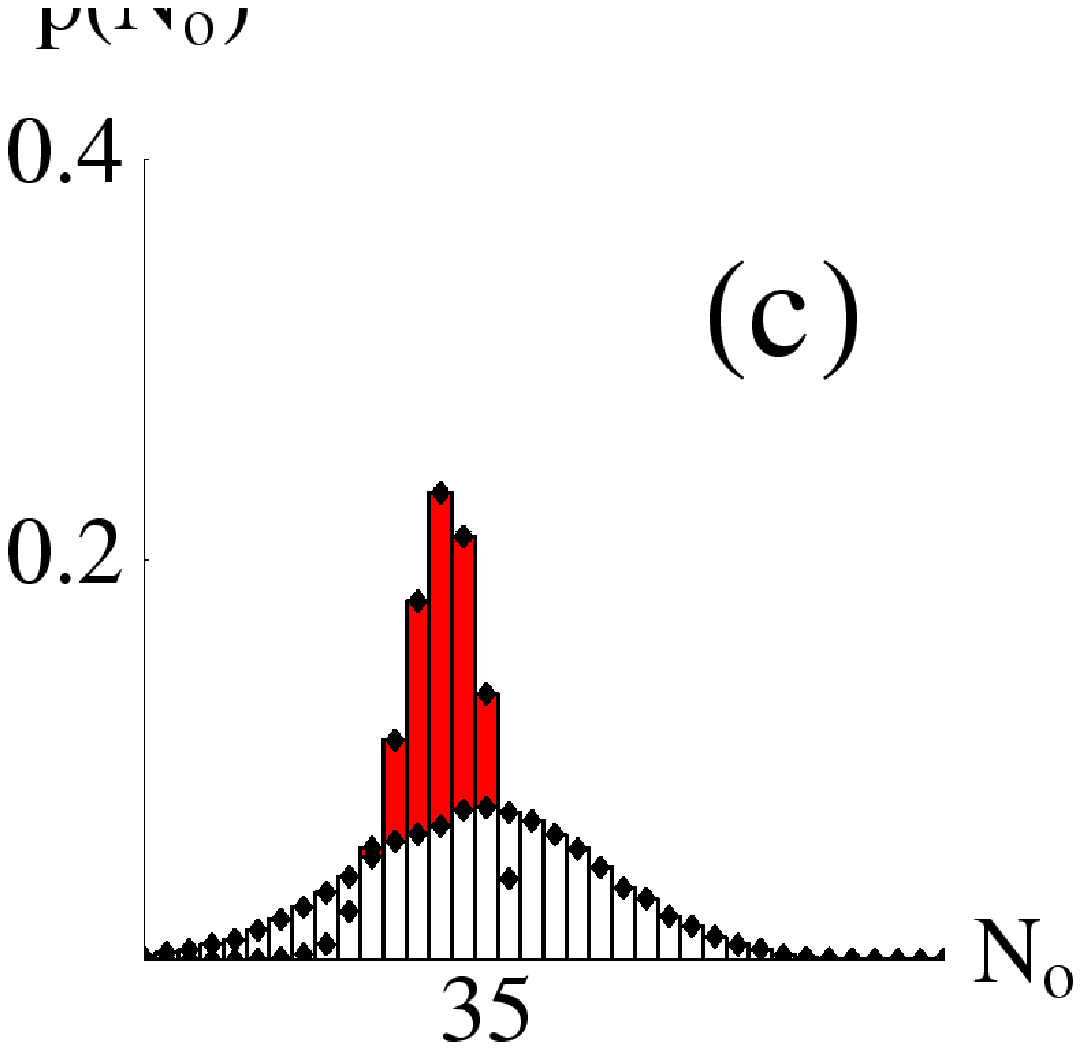}\hspace{-0.4cm}
\includegraphics[width=0.28\columnwidth]{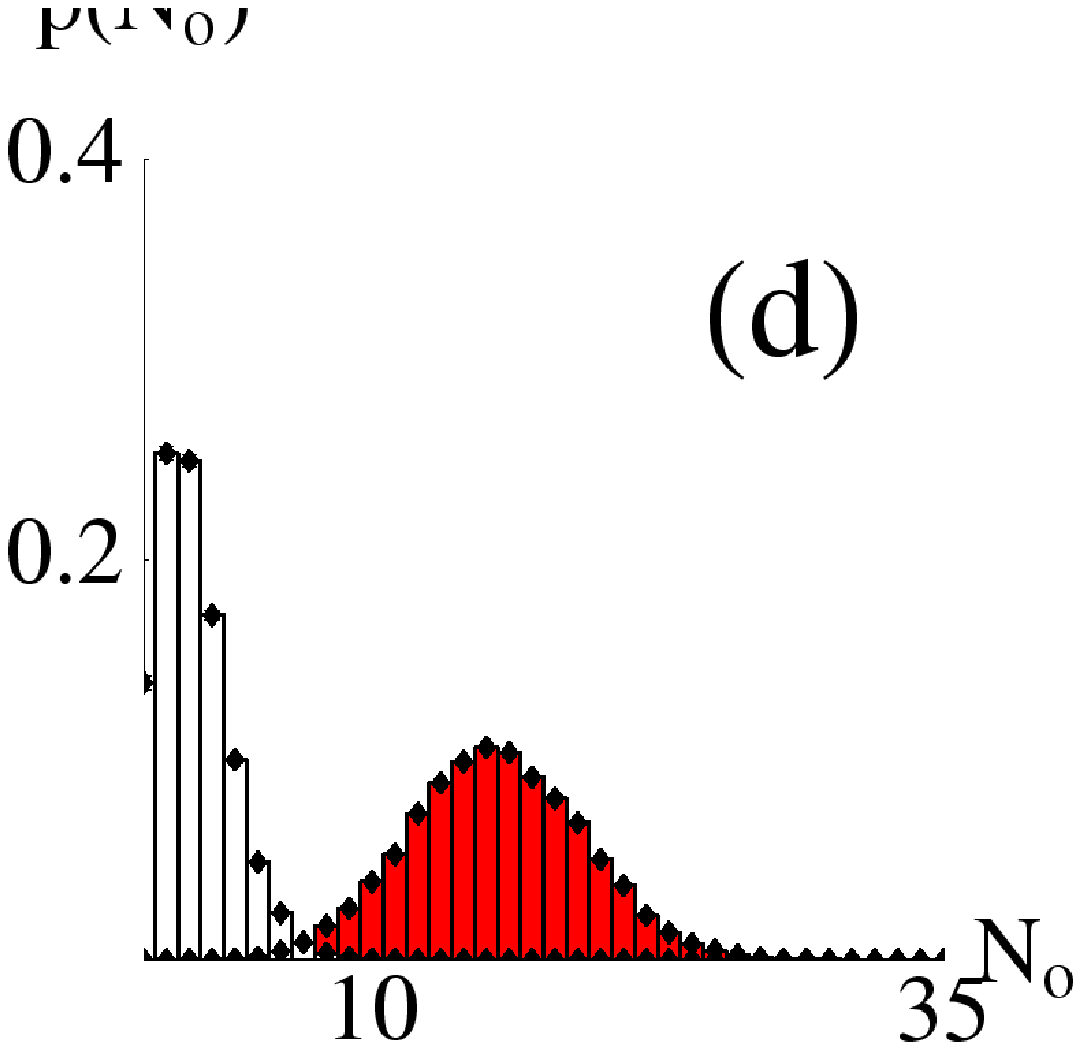}\\
\vspace{0.0cm}

\caption{(a) Results from our Monte Carlo calculation of  efficiency $\eta$ of
HLI (diamonds, $I_0=100$W/m$^2$) and LLI (crosses,
$I_0=10$W/m$^2$) membranes, as a function of incident light
intensity $I/I_0$. Continuous lines give the result of the
analytical model.  Panels (b), (c) and (d)  show the distribution
$p(N_o)$ of the number of open RCs for light intensities
corresponding to arrows in the main plot. HLI  are shaded bars and
LLI  are white bars.}\label{etaI}
\end{figure}

For completeness, we now quantify the effect of incident light
intensity variations relative to the light intensity during
growth, with both membranes having $\tau=3$ms. The externally
applied light intensity $I/I_0$, which corresponds to the ratio
between the actual ($I$) and growth ($I_0$) light intensities, is
varied in Fig. \ref{etaI}(a). The LLI membrane performance starts
to diminish well beyond the growth light intensity, while the HLI
adaptation starts diminishing just above $I_0$ due to increased
dissipation. The crossover in efficiency at $I\approx I_0$ results
from the quite different behaviors of the membranes as the light
intensity increases. In particular,  in LLI membranes excess
photons are readily used for bacterial metabolism, and HLI
membranes exploit dissipation in order to limit the number of
processed excitations.  Figs. \ref{etaI}(b), (c) and (d) verify
that performance of membranes heavily depends on the number of
open RCs. For instance, membranes subject to low excitation
intensity (Fig. \ref{etaI}(b)) behave  similarly to that expected
for fast RC cycling times  (Fig. \ref{etatau}(a)).  The complete
distributions, both for HLI and LLI conditions, shift to lower
$N_o$ with increased intensity in the same manner as that observed
with $\tau$. Even though these adaptations show such distinct
features in the experimentally relevant regimes for the RC-cycling
time and illumination intensity magnitude \cite{sturg1,RC1,RC2},
Figs.\ref{etatau}(c) and (d)  show that the distributions of open
RCs actually overlap. Despite the fact that the adaptations arise
under different  environmental conditions, the resulting dynamics
of the membranes are quite similar. Note that within this
parameter subspace of $I$ and $\tau$, the LLI membrane may have a
larger number of open RCs than the HLI adaptation. In such a case,
the LLI membrane will perform better than HLI with respect to RC
ionization. The inclusion of RC dynamics implies that the absorbed
excitation will not find all RCs available. Instead, a given
amount of closed RCs  will eventually alter the excitation's fate
since probable states of oxidization are readily reduced. In a
given lifetime, an excitation will find (depending on $\tau$ and
$I$) a number of available RCs -- which we refer to as {\it
effective stoichiometry} -- which is different from the actual
number reported by Atomic Force Microscopy
\cite{sturg1,bahatyrova}.

\begin{figure}
\begin{center}
\includegraphics[width=0.6\columnwidth]{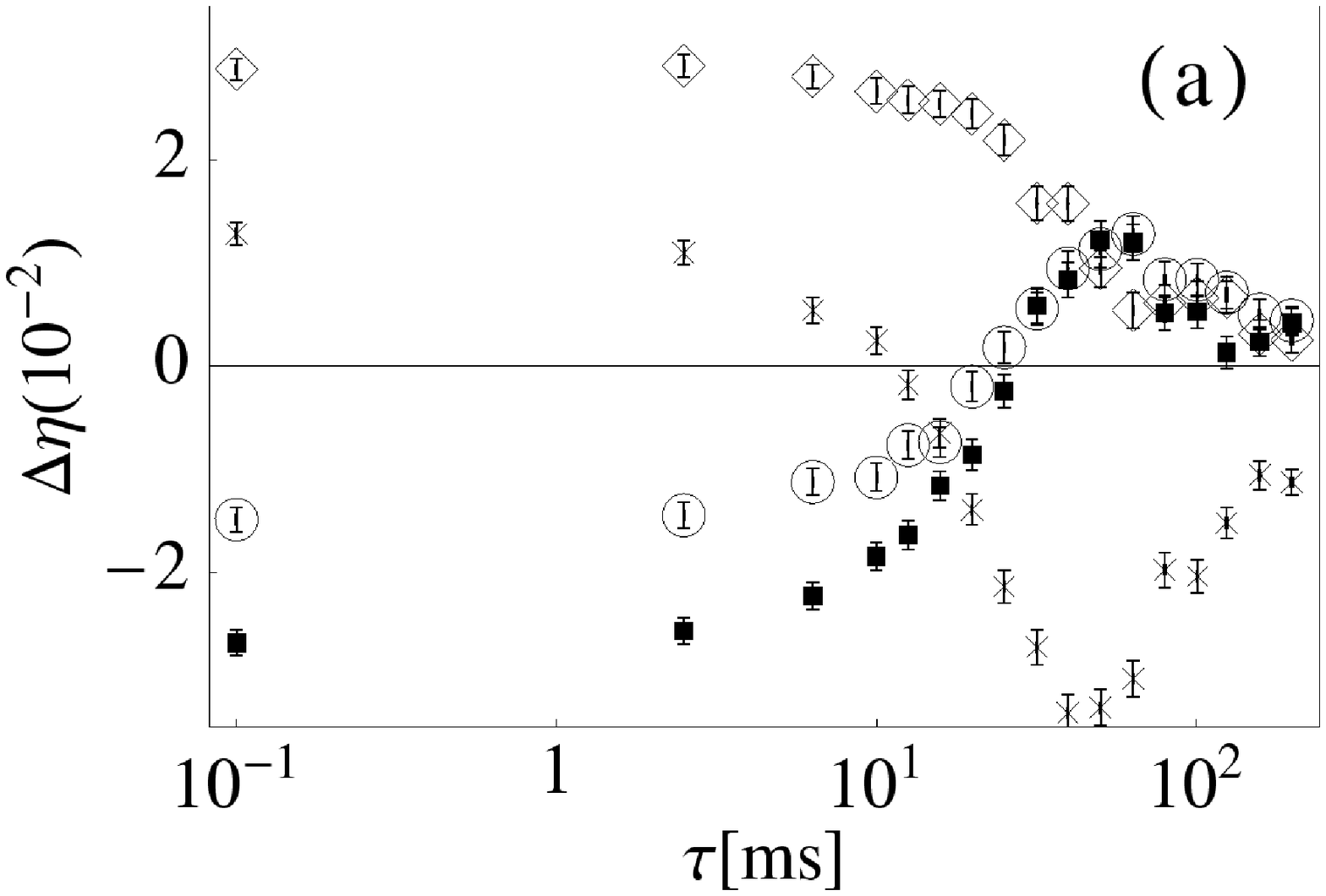}
\includegraphics[width=1.\columnwidth]{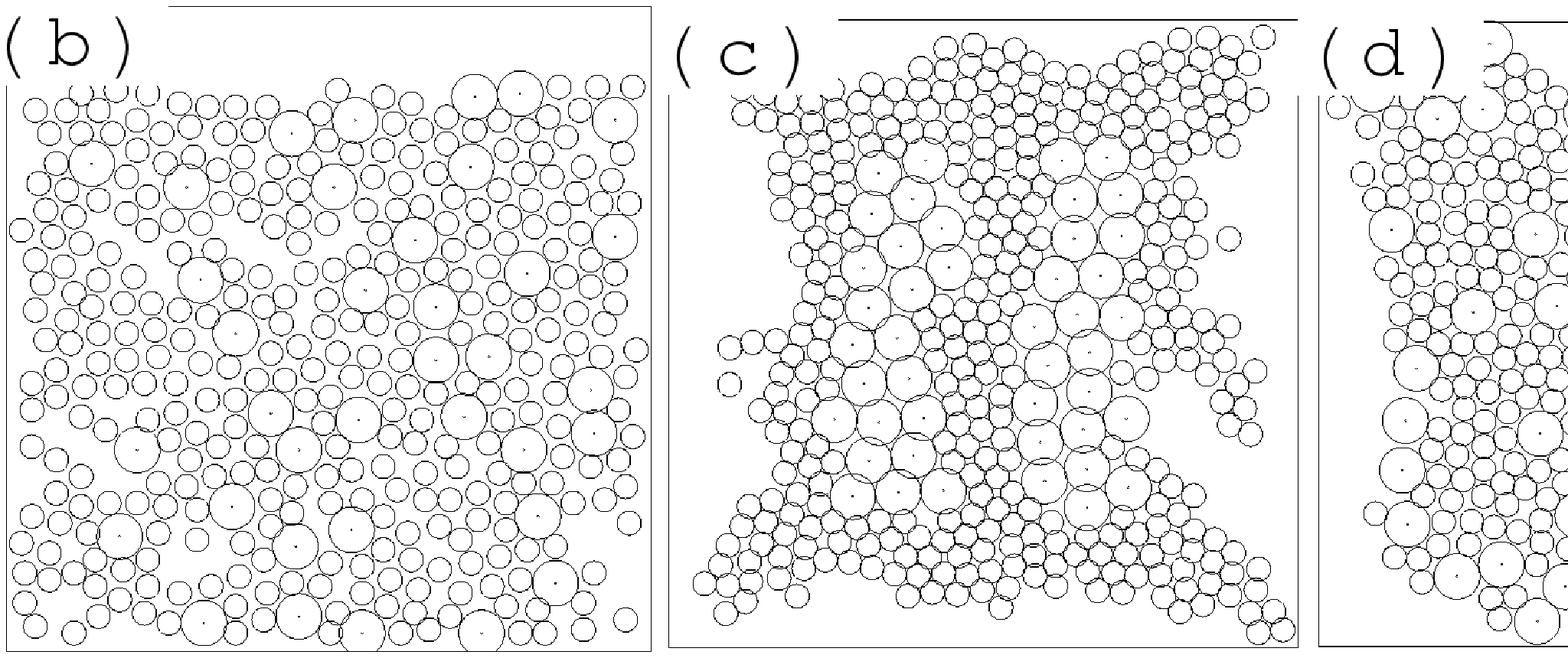}
\end{center}
\caption{(a)  $\Delta \eta$ is presented for the following membrane configurations:
(b)  {\it Rsp. Photometricum} bacteria;  (c) {\it Rs. Palustris} bacteria;
(d) completely unclustered vesicle; (e) fully clustered vesicle. The symbols are crosses, circles, diamonds and boxes,
respectively}\label{fig23.2}
\end{figure}

\section{Clustering trends}\label{archit}
The empirical AFM investigations show two main features which highlight the architectural change in the
membrane as a result of purple bacteria's  adaptation: the
 stoichiometry variation and the trend in clustering.
Fig. \ref{fig23.2} shows the importance of the arrangement of
complexes by comparing architectures (b), (c), (d) and (e), all of which
have a stoichiometry which is consistent with LLI vesicles.
Fig. \ref{fig23.2}(a) shows the difference between a given
membrane's efficiency $\eta$, and the mean of all the membranes
$\bar{\eta}$, i.e. $\Delta\eta=\eta-\bar{\eta}$. The more
clustered the RCs, the lower the efficiency in the short $\tau$
domain.  As RC cycling is increased, (b) becomes the least
efficient while all other configurations perform almost equally.
The explanation comes from the importance of the number of open
RCs: As $\tau$ gets larger, many RCs will close and the situation
becomes critical at $\tau\approx 3 $ms, where $\eta$ decreases
rapidly. Configurations (c), (d) and (e) all have the same number of RCs
(i.e. 44), and the distribution of open RCs is  almost  the
same in each case  for any fixed RC cycling time. By contrast, (b)
has fewer RCs (i.e. 36). Therefore when $\tau$ is small, sparser RCs
and exciton kinetics imply that the membrane architecture (b) will have better efficiency
than (c) and (e). The effect of the arrangement itself is lost due
to slower RC dynamics, and the figure of merit that determines
efficiency is the number of open RCs, which is lower
for (b).

To summarize so far, we find that the arrangement of complexes changes
slightly the efficiency of the membranes when no RC dynamics is
included -- but with RC dynamics, the most important feature is
the number of open RCs which is smaller for (b).
The nearly equal efficiency over the millisecond $\tau$ domain,
emphasizes the relative insensitivity to the complexes'
geometrical arrangement. The slower the RC cycling, the more
evenly available RCs will be dispersed in clustered configurations,
resembling the behavior of sparse RC membranes. Incoming
excitations in clustered configurations may quickly reach a cluster
bordering closed RCs, but must then explore further in order to
generate a charge separation. Although the longer RC closing times
make membranes more prone to dissipation and decreased efficiency,
it also makes the architecture less relevant for the overall
dynamics. {\em The relevant network architecture instead becomes the
dynamical one including sparse  open RCs, not the static
geometrical one involving the actual clustered RCs.} The inner RCs
in clusters are able to accept excitations as cycling times
increase,  and hence the  RCs overall are used more evenly. This
implies that there is little effect of the actual configuration,
and explains the closeness of efficiencies for different
arrangements in the millisecond range.

\section{Global membrane model incorporating excitation kinetics and RC-cycling}\label{analyt}

Within a typical fluorescence lifetime of 1 ns, a single
excitation has travelled hundreds of sites and explored the
available RCs globally. The actual arrangement or architecture of
the complexes seems not to influence the excitation's fate, since
the light intensity and  RC cycling determine the number of open
RCs and the availability for P oxidation. This implies that the
full numerical analysis of the excitation kinetics, while
technically more accurate, may represent excessive computational
effort -- either due to the size of the state space within the
master equation approach, or the number of runs required for
ensemble averages with the stochastic method. In addition, within
neither numerical approach is it possible to deduce the direct
functional dependence of the efficiency on the parameters
describing the underlying processes. To address these issues, we
present here an alternative rate model which is inspired by the
findings of the numerical simulations, but which (1) globally
describes the excitation dynamics and RC cycling, (2) leads to
analytical expressions for the efficiency of the membrane and the
rate of quinol production, and (3) sheds light on the trade-off
between RC-cycling and exciton dynamics \cite{superPRL}.

\begin{figure}
\centering
\includegraphics[width=0.45\columnwidth]{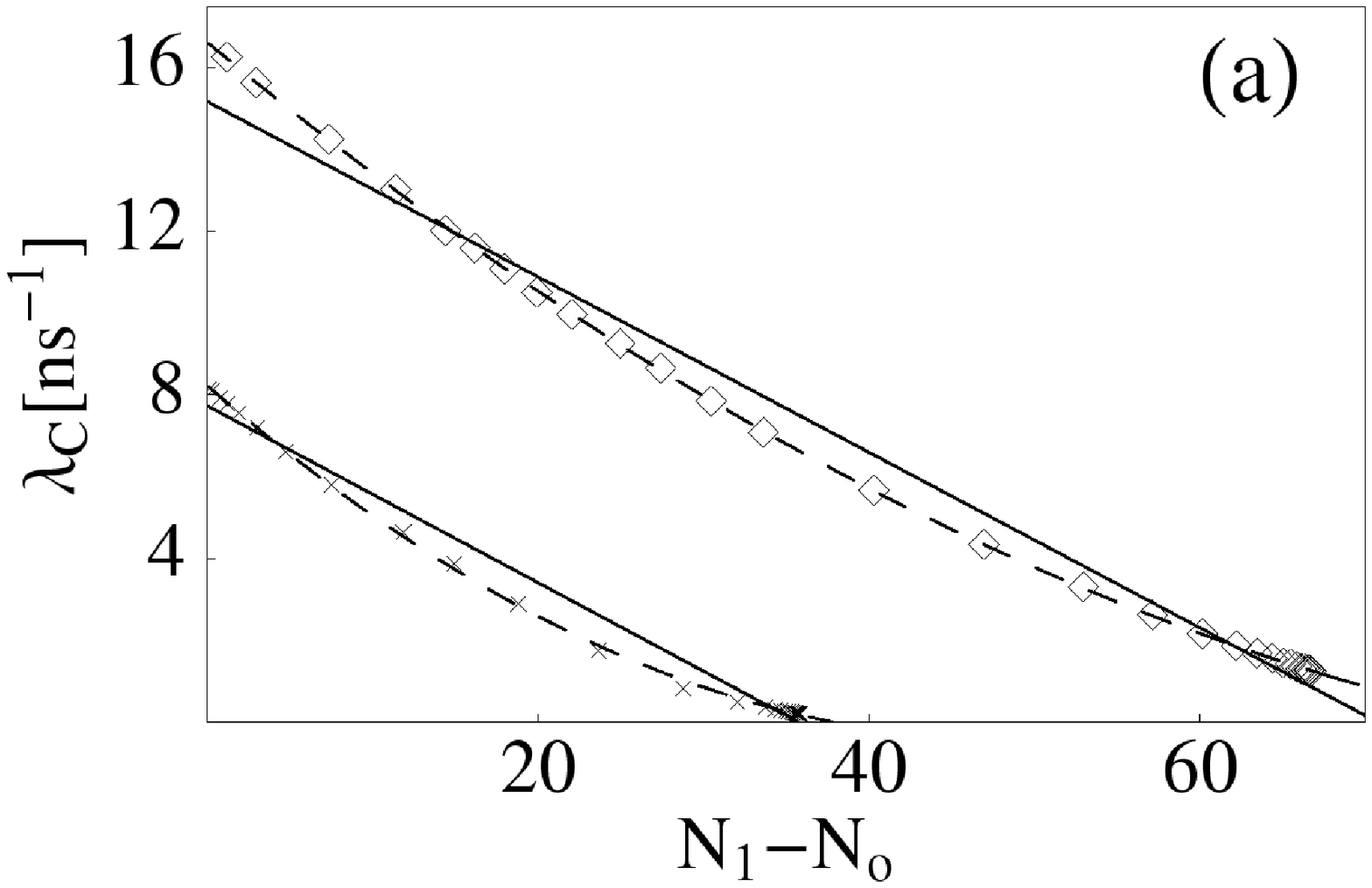}
\includegraphics[width=0.45\columnwidth]{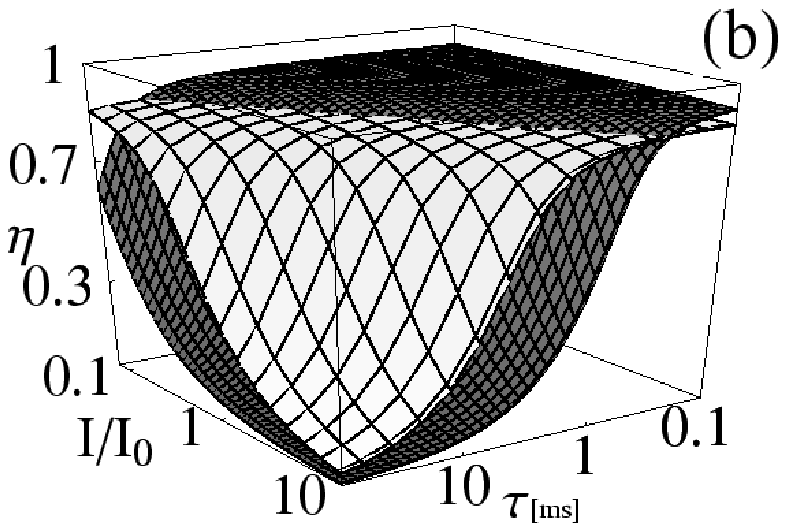}\\
\caption{(a) Numerical results showing the rate of ionization
$\lambda_C(N_o)$ of an RC for HLI (diamonds) and LLI (crosses)
membranes, together with a quadratic (dashed line) and linear
(continuous) dependence on the number of closed RCs
$(N_{1}-N_{o})$. The fitting parameters for $a+bN_o$ are
$a=\{15.16, 7.72\} $ns$^{-1}$, $b=\{-0.21,-0.21\}$ ns$^{-1}$; and
for  $a+bN_o+cN_o^2$, $a=\{16.61,8.21\}$ns$^{-1}$,
$b=\{-0.35,-0.33\}$, and $c=\{3.6, 1.5\} \mu$s$^{-1}$, for HLI and
LLI membranes respectively. (b) $\eta$ as function of $\tau$ and
$\alpha=I/I_0$, obtained from the complete analytical solution for
LLI (white) and HLI (grey) membranes}\label{eta3D}
\end{figure}

We start with the observation that absorbed excitations are
transferred to RCs, and finally ionize special pairs or are
dissipated. At any given time $N_E$ excitations will be present in
the membrane. The rate at which they are absorbed is $\gamma_{A}$.
Excitations reduce quinone $Q_B$ in the membrane at RCs due to P
oxidation at a rate $\lambda_C N_E$,  or dissipate at a rate
$\gamma_D N_E$. Both processes imply that excitations leave the
membrane at a rate $\frac{dN_E}{dt}=-\lambda_C N_E-\gamma_D N_E$.
Here $\lambda_C$ is the inverse of the mean time $\tau_{RC}$ at
which an excitation yields a charge separation at RCs when
starting from any given complex, and it depends on the current
number of open RCs given by $N_{o}$. The RC cycling dynamics
depend on the rate at which  RCs close, i.e.
$-\frac{\lambda_C(N_{o})}{2} N_E$ where the 1/2 factor accounts
for the need for two excitations to produce quinol and close the
RC. The  RCs open at a rate $1/\tau$, proportional to the current
number of closed RCs given by $N_{1}-N_{o}$. Hence the
RC-excitation dynamics can be represented by two nonlinear coupled
differential equations:
\begin{eqnarray}
\frac{dN_E}{dt}&=&-(\lambda_C(N_{o})+\gamma_D) N_E+\gamma_{A}  \label{ne}\\
\frac{dN_{o}}{dt}&=&\frac{1}{\tau}(N_{1}-N_{o})-\frac{\lambda_{C}(N_o)}{2} N_E. \label{noff}
\end{eqnarray}
In the stationary state, the number of absorbed excitations $n_A$
in a time interval $\Delta t$ and the number of excitations used
to produce quinol $n_{RC}$, are given by
\begin{eqnarray}
n_A&=& \gamma_A \Delta t\\
n_{RC}&=&\lambda_C(N_{o}) N_E \Delta t
\end{eqnarray}
yielding an expression for the steady-state efficiency $\eta=n_{RC}/n_A$:
\begin{eqnarray}
\eta=\frac{\lambda_C(N_{o}) N_E}{\gamma_A}.
\end{eqnarray}
These equations can be solved in the stationary state for $N_E$
and $N_{o}$,  algebraically or numerically,  if the functional
dependence of $\lambda_C(N_{o})$ is given.  It is zero when all
RCs are closed, and a maximum $\lambda_C^0$  when all are open.
Making the functional dependence explicit, $\lambda_C(N_{o})$,
Fig. \ref{eta3D}(a) presents the relevant functional form for HLI
and LLI membranes together with a linear and a quadratic fit. The
dependence on the rate of quinone reduction  $\lambda_C(N_{o})$
requires quantification of the number of open RCs, with a notation
where the fitting parameter comprises duplets where first and
second components relate to the HLI and LLI membranes being
studied. Figure \ref{eta3D}(a) shows that $\lambda_C(N_{o})$
favors a quadratic dependence of the form $\lambda_C(N_{o})= a+b
N_{o}+c N_{o}^2.$ A linear fit
\begin{eqnarray}
\lambda_C(N_{o})=\lambda_C^0\left( \frac{N_{o}}{N_{1}}\right)\label{linear}
\end{eqnarray}
smears out the apparent power-law behavior with fit value
$N_{1}=\{70.72, 35.71\}$, in close agreement with the HLI and LLI
membranes which have 67 and 36 RCs respectively. The linear fit
$\lambda_C(N_o)$  can used to generate  an analytical expression
for $\eta$ as follows:
\begin{eqnarray}
 \eta(\tau,\gamma_A(I))=\frac{1}{\gamma_A\lambda_C^0\tau}\left\{2N_1(\lambda_C^0+\gamma_D)+\gamma_A\lambda_C^0\tau-\right.& &\\
\left.\sqrt{4N_{1}^2(\lambda_C^0+\gamma_D)^2+4 N_{1}\gamma_A\lambda_C^0(\gamma_D-\lambda_C^0)\tau+(\gamma_A\lambda_C^0\tau)^2}\right\}\ \ . \nonumber
\end{eqnarray}
We found no analytical solution for $\eta$ in the case where
$\lambda_C(N_{o})$ has a power law dependence. In the limit of
fast RC cycling-time ($\tau$$\rightarrow$0), $\eta$ has the simple
form $\eta=(1+\gamma_D/\lambda_C^0)^{-1}$. If all transfer paths
are summarized by $\lambda_C^0$, this solution illustrates that
$\eta \geq 0.9$ if the transfer-P reduction time  is less than a
tenth of the dissipation time, not including RC cycling. As can be
seen in  Figs. \ref{etatau} and \ref{etaI}, the analytical
solution is in good   quantitative agreement with the numerical
stochastic simulation, and provides support for the assumptions
that we have made. Moreover, our theory shows directly that the
efficiency is driven by the interplay between the RC cycling time
and light intensity. Figure \ref{eta3D}(b) shows up an entire
region of parameter space where LLI membranes are better than HLI
in terms of causing P ionization, even though the actual number of
RCs that they have is smaller. In view of these results, it is
interesting to note how clever Nature has been in tinkering with
the efficiency of LLI vesicles and the dissipative behavior of HLI
adaptation, in order to meet the needs of bacteria subject to the
illumination conditions of the growing environment.

\section{Bacterial metabolic demands: A stoichiometry prediction from our model}\label{metab}

\begin{figure}
\includegraphics[width=0.47\columnwidth]{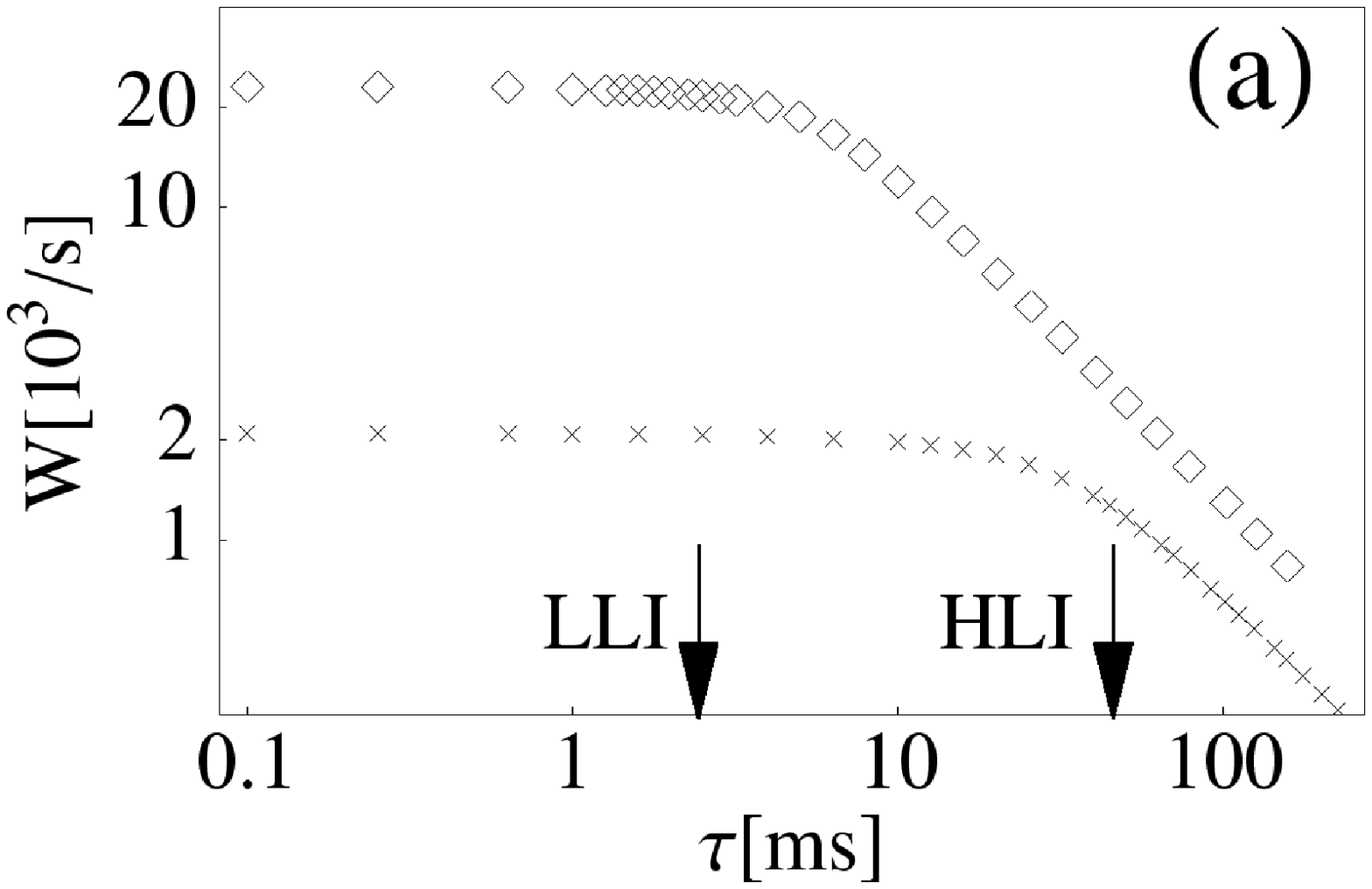}
\includegraphics[width=0.47\columnwidth]{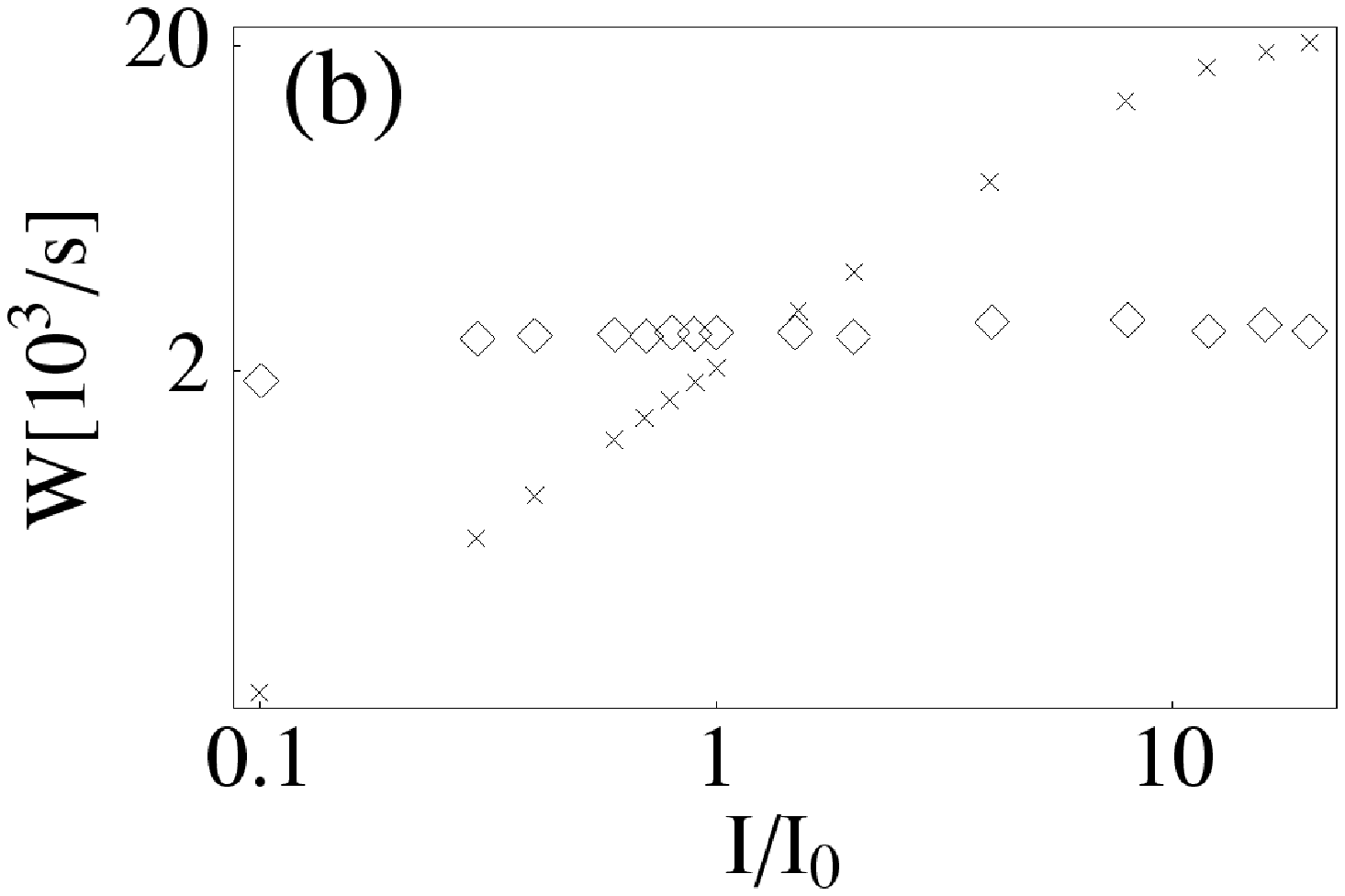}
\caption{ (a) Quinol rate $W$ in HLI (diamonds, $I_0=100$W/m$^2$)
and LLI (crosses, $I_0=10$W/m$^2$) grown membranes, as a function
of RC cycling time. The times shown with arrows are used  in (b),
where $W$ is presented as a function of incident
intensity.}\label{metW}
\end{figure}

Photosynthetic membranes must provide enough energy to fulfill the
metabolic requirements of the living bacteria. In order to quantify the
quinol output of the vesicle, we calculate the quinol rate
\begin{equation}
W=\frac{1}{2}\frac{dn_{RC}}{dt}
\end{equation}
which depends directly on the excitations that ionize RCs $n_{RC}$.
The factor $\frac 12$ accounts for the requirement of two
ionizations to form a single quinol molecule. Fig. \ref{metW}(a)
shows the quinol rate as a function of RC cycling time, when
membranes are illuminated in their respective grown conditions. If
RC cycling is not included ($\tau\rightarrow 0$) the tenfold
quinol output difference suggests that the HLI membrane could
increase the cytoplasmic pH to dangerously high levels, or that the LLI
membrane could starve the bacteria. However,  the bacteria manage
to survive in both these conditions -- and below we explain why.

In the regime of millisecond
RC cycling, the quinol rate in HLI conditions decreases which is
explained by dissipation enhancement acquired from only very few
open RCs. Such behavior in LLI conditions appears only after
several tens of milliseconds. The fact that no crossover occurs in
quinol rate for these two  membranes, suggests that different
cycling times generate this effect. The arrows in Fig.
\ref{metW}(a) correspond to times where a similar quinol rate is
produced in both membranes, in complete accordance with numerical
studies where enhanced quinone diffusion lessens RC cycling times
\cite{sturgdiff}  in LLI adaptation. Although these membranes were
grown under continuous illumination, the adaptations themselves
are a product of millions of years of evolution. Using RC cycling
times that preserve quinol rate in both adaptations, different
behaviors emerge when the illumination intensity is varied (see
Fig. \ref{metW}). The increased illumination is readily  used by
the LLI adaptation, in order to profit from excess excitations in
an otherwise low productivity regime. On the other hand, the HLI
membrane maintains the quinone rate constant, thereby avoiding the
risk of pH imbalance in the event that the light intensity
suddenly increased. We stress that the number of RCs synthesized does not
directly reflect the number of available states of ionization in
the membrane. LLI synthesizes a small amount of RCs in order
to enhance quinone diffusion, such that excess light intensity is
utilized by the majority of special pairs. In HLI, the synthesis
of  more LH1-RC complexes  slows down RC-cycling, which ensures
that many of these RCs are unavailable  and hence a steady quinol
supply is maintained independent of any excitation increase. The
very good agreement between our analytic results and the stochastic
simulations, yields additional physical insight concerning the
stoichiometries found experimentally in {\it Rsp.
Photometricum}\cite{sturg1}. In particular, the vesicles studied
repeatedly exhibit the same stoichiometries, $s\approx 4$ for HLI,
and $s\approx 8$ for LLI membranes. Interestingly, neither smaller nor
intermediate values are found.

\begin{figure}
\centering
\includegraphics[width=0.48\columnwidth]{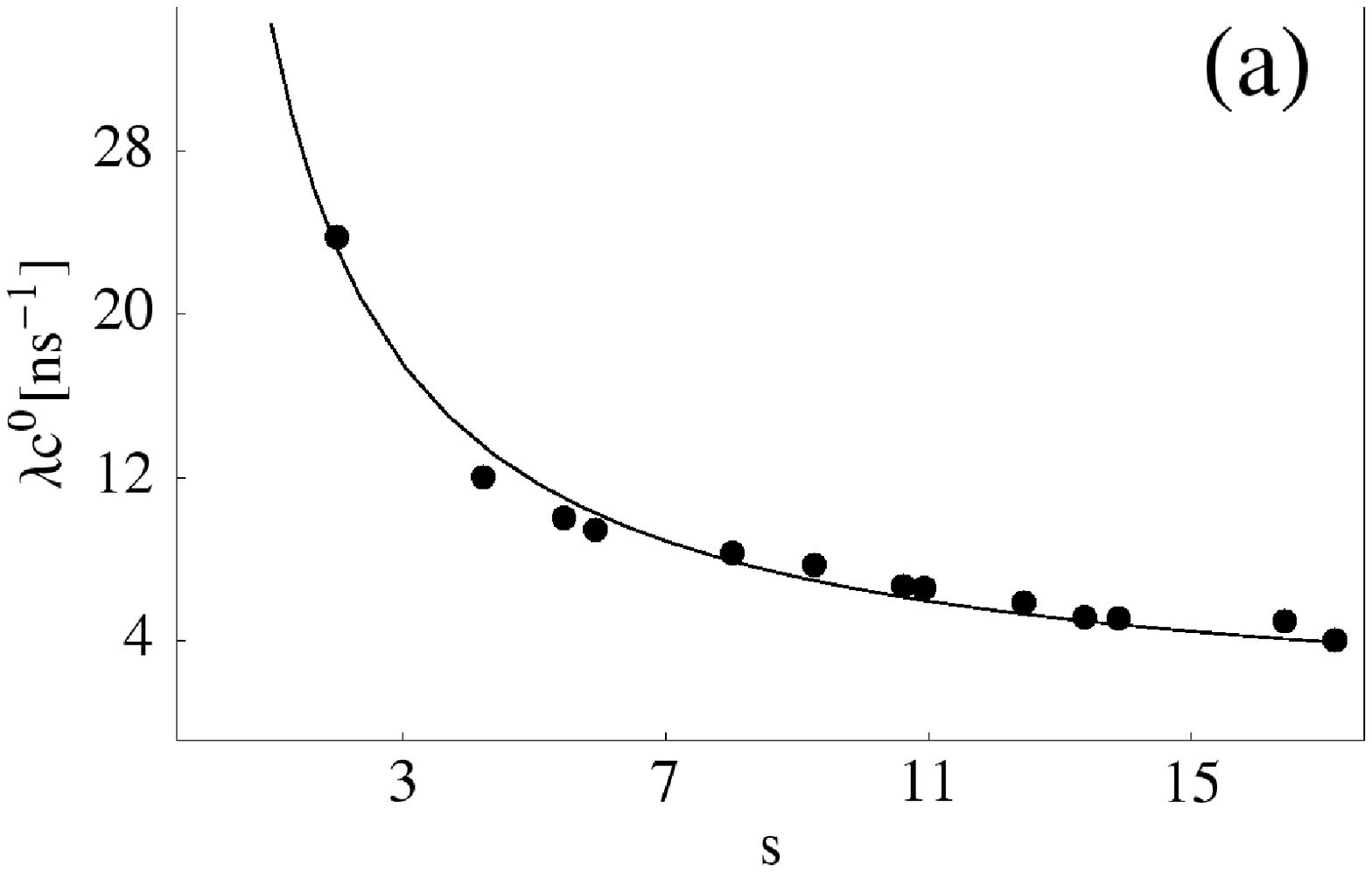}
\includegraphics[width=0.43\columnwidth]{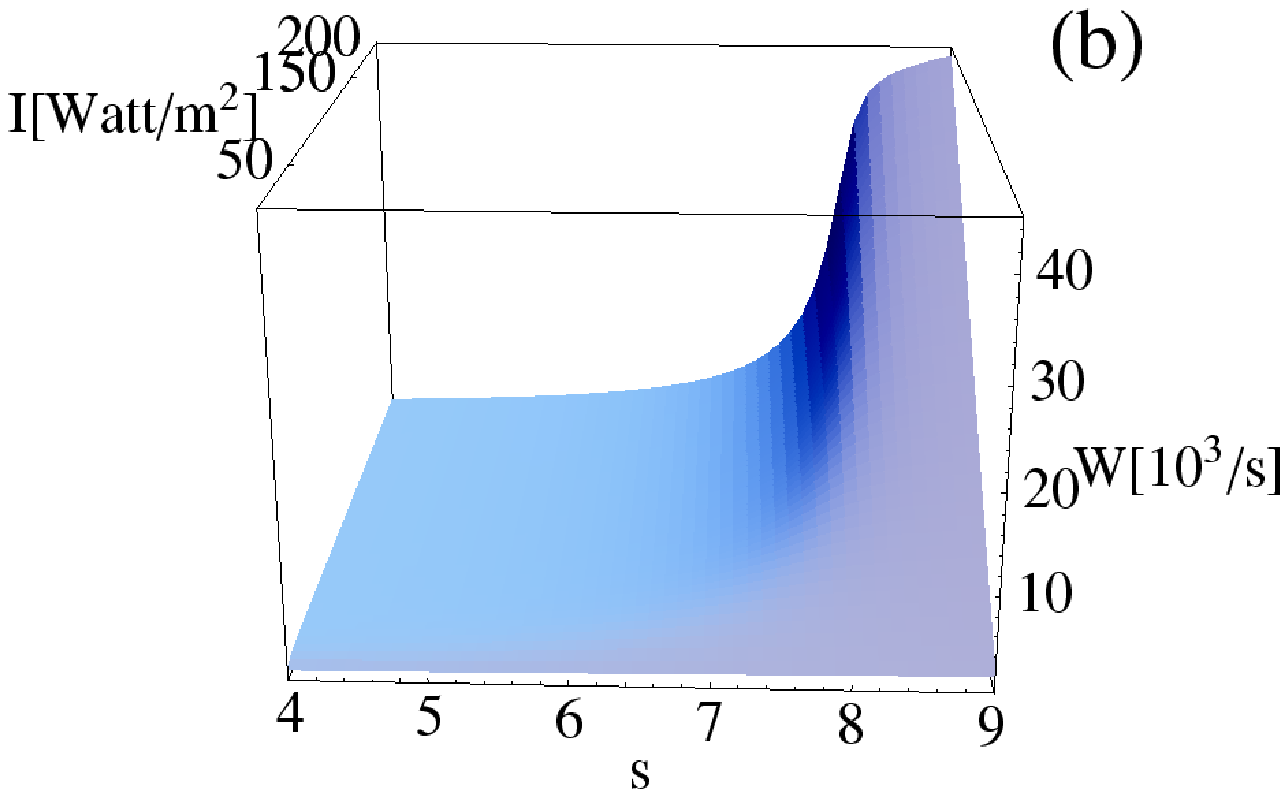}
\includegraphics[width=0.5\columnwidth]{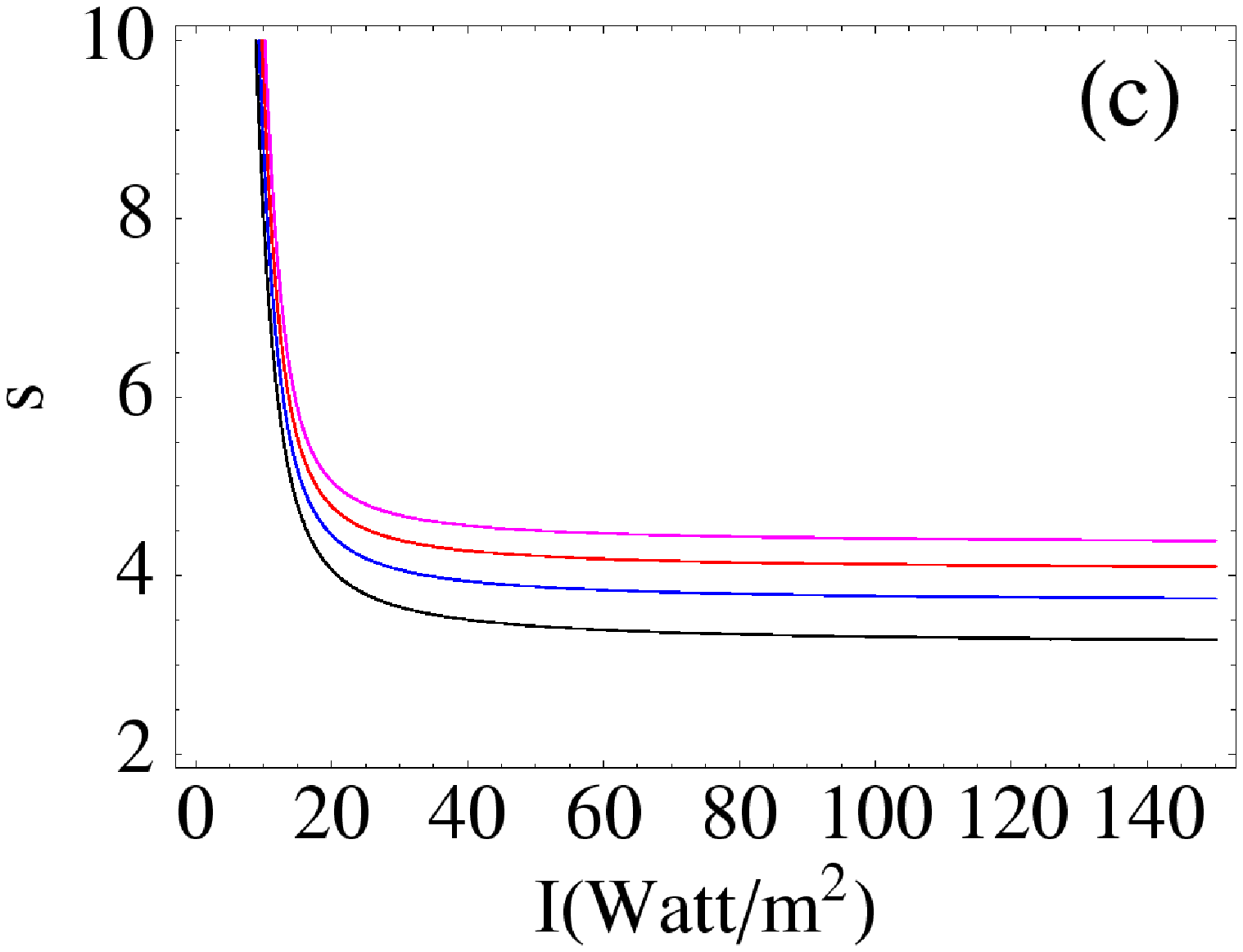}
\caption{ (a) Dependance of $\lambda_C^0$ on stoichiometry $s$ of the membrane.  (b) $W(s,I)$ as a function
of stoichiometry $s$ variation and illumination intensity.  (c)
Quinol rate contours of $W=\{1900,2000,2100,2200\}$ s$^{-1}$ in black, blue, red and pink, respectively.}\label{spred}
\end{figure}

We now derive an approximate expression for the quinol production
rate $W$ in terms of the environmental growth conditions and the
responsiveness of purple bacteria through stoichiometry
adaptation. Following Refs.
\cite{sturg1,sturgdiff,scheuring2004b}, the area $A_0$ of
chromatophores in different light intensities can be assumed
comparable. Initially absorption occurs with $N_1$ LH1 complexes
of area $A_1$, and $N_2$ LH2 complexes of area $A_2$, which fill a
fraction $f$ of the total vesicle area $f=(A_1N_1+A_2N_2)/A_0$.
This surface occupancy can be rearranged in terms of the number of
RCs $N_1$, and the stoichiometry $N_1=N_2/s$, yielding  the
expression $f=N_{1}(A_1+sA_2)/A_0$. The fraction $f$  has been
shown \cite{sturgdiff} to vary among adaptations $f(s)$, since LLI
have a greater occupancy than HLI membranes due to
para-crystalline LH2 domains in LLI. Accordingly,  the absorption
rate can be cast as $\gamma_A=I(\gamma_1+s\gamma_2)\frac{f(s)
A_0}{A_1+s A_2}$. Following photon absorption, the quinol
production rate $W=\lambda_C N_E/2$ depends on the number of
excitations within the membrane in the stationary state, and on
the details of transfer  through the rate $\lambda_C$. The assumed
linear dependence $\lambda_C(N_o)$ requires knowledge of
$\lambda_C^0$ and the stationary-state solution for the mean
number of closed RCs through
$N_{o}=N_1-\frac{\lambda_C\lambda_A}{2(\gamma_D+\lambda_C)}\tau$.
The stationary state in Eq. \ref{ne} yields $
N_E=\frac{\lambda_A}{\gamma_D+\lambda_C} $ such that the mean
number of closed RCs is simply $ N_o =N_1-W\tau $. The rate
$\lambda_C^0$ is the rate at which excitations oxidize any special
pair when all RCs are open. The time to reach an RC essentially
depends on the number of RCs and hence the stoichiometry $s$.
$\lambda_C^0$  must be zero when no RCs are present
($s$$\rightarrow$$\infty$) and takes a given value $\langle
t_0\rangle^{-1}$ when the membrane is made of only LH1s ($s$=0).
Also the RC cycling time $\tau$ is expected to vary somewhat with
adaptations due to quinone diffusion \cite{sturgdiff},  which is
supported in our analysis by the condition of bounded  metabolic
demands as presented in Fig. \ref{metW}(a). The linear assumed
dependence of Eq. \ref{linear} is cast explicitly as a function of
$W$ and $s$ with the  number of open RCs
\begin{equation}
\lambda_C(s,W)=\lambda_C^0(s)\left(1-\frac{W\tau(s)(A_2s+A_1)}{A\ f(s)}\right)\ .
\end{equation}
From $N_E$ in the steady state, we have
\begin{equation}
W=\frac{\lambda_C(s,W)\lambda_A(s,I)}{2(\lambda_C(s,W)+\gamma_D)},
\end{equation}
which can be solved for $W(s,I)$
\noindent{\small{\small\begin{equation}
2W(s,I)=\frac{\gamma_A(s,I)}{2}+ \frac{1}{B(s)}
\left(1+\frac{\gamma_{D}}{\lambda_c^0}\right)
+
\sqrt{[\frac{\gamma_A(s,I)}{2}+\frac{1}{B(s)}\left(1+\frac{\gamma_{D}}{\lambda_c^0}\right)]^2+\frac{\gamma_A(s,I)}{2B(s)}}
\end{equation}}}
\noindent where $B(s)=\frac{\tau(s) (A_1 +s A_2)}{f(s)A_0}$. To
determine $\tau(s)$, we employ a linear interpolation using the
values highlighted by arrows in Fig. \ref{metW}(b). The
requirements on  $\lambda_C^0(s)$ are fulfilled by a fit
$\lambda_C^0(s)=(s/a+\langle t_0\rangle)^{-1}$, which is supported
by our calculations for configurations of different stoichiometries
(see Fig. \ref{spred}(a)). The filling fraction fraction $f(s)$ is
assumed linear according to  the experimentally found values for
HLI  ($f\approx 0.75$) and LLI  ($f\approx 0.85$). The resulting
expression is presented in  Fig. \ref{spred}(b). In the high
stoichiometry/high intensity regime, the high excitation number
would dangerously increase  the cytoplasmic pH
\cite{fassioli,Geyer,review}. The contours shown in Fig.
\ref{spred}(c) of constant quinol production rate $W$, show that
only in a very small intensity range will bacteria adapt with
stoichiometries which are different from those experimentally
observed in {\it Rsp. Photometricum} ($s\approx $4 and
$s\approx$8). As emphasized in Ref.  \cite{sturg1},  membranes
with $s$=6 or $s$=2 were not observed, which is consistent with
our model. More generally, our results predict a great sensitivity
of stoichiometry ratios for 30-40 W/m$^2$, below which membranes
rapidly build up the number of antenna LH2 complexes. Very
recently \cite{sturg2009}, membranes were grown with 30W/m$^2$ and
an experimental stoichiometry of 4.8 was found. The contour of
2200 s$^{-1}$ predicts a value for stoichiometry of 4.72 at such
light intensities. This agreement is quite remarkable, since a
simple linear model would   wrongly predict $s=7.1$. Our theory's
full range of predicted behaviors as a function of light-intensity
and stoichiometry, awaits future experimental verification.

\section{Concluding remarks}
We have shown that excitation dynamics alone cannot explain the
empirically observed adaptation of light-harvesting membranes.
Instead, we have presented a quantitative model which strongly
suggests that chromatic adaptation results from the interplay
between excitation kinetics and RC charge carrier quinone-quinol
dynamics. Specifically, the trade-off between light intensity and
RC cycling dynamics induces LLI  adaptation in order to
efficiently promote P oxidation due to the high amount of open
RCs. By contrast, the HLI membrane remains less efficient in order
to provide the bacteria with a relatively steady metabolic
nutrient supply.

This successful demonstration of the interplay between excitation
transfer and  RC trapping, highlights the important middle ground
which photosynthesis seems to occupy between the fast dynamical
excitation regime in which quantum-mechanical calculations are
most relevant \cite{fassioli,review,plenio}, and the purely
classical regime of the millisecond timescale bottleneck in
complete membranes. We hope our work will encourage further study
of the implications for photosynthesis of this fascinating
transition regime between quantum and classical
behaviors\cite{plenio2}. On a more practical level, we hope that
our study may help guide the design of more efficient solar
micropanels mimicking natural designs.
\section*{Aknowledgements}
This research was supported by Banco de la Rep\'ublica (Colombia)
and Proyecto Semilla  (2010-2011), Facultad de Ciencias
Universidad de los Andes.
\\
\appendix
\section*{Appendix}
\paragraph{Residence time $t_{H_k}$}
In the stochastic simulations, $t_i^a$ is the residence time of an
excitation in the $a^{th}$ realization at complex $i$:
\begin{equation}
\hat{t}_{k}=\frac{\sum_{a, i\epsilon k} t_i^a}{\sum_{i\epsilon k} n_{V_i}}
\end{equation}
where $n_{V_i}$ is the number of times complex $i$ has been
visited in all the stochastic realizations.

\paragraph{Dissipation $d_i$}
The dissipation $d_i$ measures the probability for  excitations to
dissipate at site $i$, and can be obtained formally from
\begin{equation}
d_i=\frac{n_{D_i}}{n_A}
\end{equation}
where $n_{D_i}$ are the number of excitations dissipated at site
$i$ and $n_A$ is the total number of absorbed excitations.

\paragraph{Residence probability $p_{R_k}$}
In practice within the stochastic simulations, for a given realization $a$, an
 excitation will be a time $t_1^a$ in LH1s, a time $t_2^a$ in LH2s and a time $t_3^a$ in RCs.
 The residence probability will become
\begin{equation}
p_{R_k}=\frac{\sum_a t^a_k}{\sum_{j,a} t_k^a}.
\end{equation}
The total time of all realizations in complex type $k$ is the numerator,
while the denominator stands for the total time during which the excitations were within the membrane.

\end{document}